\documentclass[aps, twocolumn, notitlepage,longbibliography, superscriptaddress, prb]{revtex4-1}

\usepackage[english]{babel}
\usepackage{mathtools}
\usepackage{graphicx}
\usepackage{epsfig}
\usepackage{wasysym}
\usepackage{amsfonts}
\usepackage{bm}
\usepackage{color}
\usepackage[normalem]{ulem}
\usepackage{cancel}
\usepackage{enumitem}

\usepackage{microtype}
\usepackage{braket}
\usepackage{multirow}

 \usepackage[hidelinks]{hyperref} 
 \hypersetup{
     colorlinks = true,
     linkcolor = blue,
     anchorcolor = blue,
     citecolor = blue,
     filecolor = blue,
     urlcolor = blue
     }

\usepackage{dsfont}

\graphicspath{{figs/}}

\newcommand{\be}{\boldsymbol{\eta}}
\newcommand{\bs}{\boldsymbol{S}}
\newcommand{\bn}{\boldsymbol{n}}
\newcommand{\ve}[1]{\boldsymbol{#1}}

\newcommand{\revisionchange}[1]{#1}

\begin{document}
	\title{Spin crossovers and {superdiffusion} in the one-dimensional Hubbard model}
	\date{\today}
\author{Michele Fava}
\affiliation{Rudolf Peierls Centre for Theoretical Physics,  Clarendon Laboratory, Oxford OX1 3PU, UK}
\author{Brayden Ware}
\affiliation{Department of Physics, University of Massachusetts, Amherst, MA 01003, USA}
\author{Sarang Gopalakrishnan}
\affiliation{Department of Physics and Astronomy, CUNY College of Staten Island,
Staten Island, NY 10314}
\affiliation{Physics Program and Initiative for the Theoretical Sciences,
The Graduate Center, CUNY, New York, NY 10016, USA}
\author{Romain Vasseur}
\affiliation{Department of Physics, University of Massachusetts, Amherst, MA 01003, USA}
\author{S. A. Parameswaran}
\affiliation{Rudolf Peierls Centre for Theoretical Physics,  Clarendon Laboratory, Oxford OX1 3PU, UK}

\begin{abstract}
We use tools from integrability and generalized hydrodynamics to study finite-temperature dynamics in the one-dimensional Hubbard model. First, we examine charge, spin, and energy transport away from half-filling and zero magnetization, focusing on the strong coupling regime where we identify a rich interplay of temperature and energy scales, with crossovers between distinct dynamical regimes. We identify an intermediate-temperature regime analogous to the spin-incoherent Luttinger liquid, where spin degrees of freedom are hot but charge degrees of freedom are at low temperature. We demonstrate that the spin Drude weight exhibits sharp features at the crossover between this regime and the low-temperature Luttinger liquid regime, that are absent in the charge and energy response, and rationalize this behavior in terms of the properties of Bethe ansatz quasiparticles. We then turn to the dynamics along special lines in the phase diagram corresponding to half-filling and/or zero magnetization where on general grounds we anticipate that the transport is sub-ballistic but superdiffusive. We provide analytical and numerical evidence for Kardar-Parisi-Zhang (KPZ) dynamical scaling (with length and time scales related via $x\sim t^{2/3}$) along both lines and at the $SO(4)$-symmetric point where they intersect. Our results suggest that both spin-coherence crossovers and KPZ scaling may be accessed in near-term experiments with optical lattice Hubbard emulators.
\end{abstract}

	\maketitle
	
	\section{\label{sec:intro}Introduction}

The Hubbard model has a storied history in the study of strong correlations  in  many-body quantum systems. Originally formulated to model interacting  electrons in narrow energy bands,~\cite{Hubb1} it came to renewed prominence following the discovery of the copper-oxide high-temperature (high-$T_c$) superconductors. The ability of the Hubbard model to capture what are believed to be key features of the high-$T_c$ phase diagram~\cite{RevModPhys.78.17} --- for example, the existence of an antiferromagnetic Mott insulator at half filling that \revisionchange{could yield}  unconventional superconducting states upon doping  
--- have made it an enduring subject of theoretical studies,~\cite{gebhard1997metal} and a favored testbed for new techniques. Numerical approaches~\cite{PhysRevX.5.041041,PhysRevX.10.031016} such as dynamical mean-field theory~\cite{RevModPhys.68.13} and density-matrix renormalization group,~\cite{Schollwock:2011aa} as well as theoretical frameworks such as quantum spin liquids~\cite{Savary_2016,QSLReview2020} and quantum criticality,~\cite{sachdev_2011} were either devised for, or greatly stimulated by, application to the Hubbard model.

An influential line of inquiry pertains specifically to the Hubbard model in {\it one} spatial dimension, which admits an exact solution via the technique of the Bethe ansatz.~\cite{essler_2005} This integrability has meant that many subtle features of the model, including non-perturbative effects, can be explored with analytical control --- including those, such as the existence of hidden symmetries,~\cite{PhysRevLett.63.2144} that extend also to its higher-dimensional counterparts. The one dimensional model can also be experimentally relevant in its own right: for instance,  one-dimensional (extended) Hubbard models have been used to describe correlations in  carbon nanotubes,~\cite{PhysRevB.55.R11973} and as a starting point for the description of materials, such as organic charge-transfer salts,~\cite{Jerome:2004aa} that can be approximated as quasi-one-dimensional.~\cite{Giamarchi:2004aa}

More recently, the Hubbard model has also  received much attention in a  setting quite distinct from its solid-state origins: namely, that of ultracold atomic gases.~\cite{JAKSCH200552,Bloch2005,doi:10.1080/00018730701223200}  Over the past two decades, 
it has been the subject of a concerted experimental effort to build `optical lattice emulators': systems of cold trapped neutral gases moving in lattice potentials and subject to strong contact interactions. The overarching goal is to engineer artificial systems whose microscopic Hamiltonian is {\it exactly} that of the Hubbard model, so as to experimentally address and potentially settle the many questions that remain the subject of spirited theoretical debate. This program has had striking successes, such as experimental realization of the {\it bosonic} Hubbard model~\cite{PhysRev.129.959} and its Mott insulator-superfluid transition~\cite{PhysRevB.34.3136,PhysRevB.37.325,PhysRevB.40.546, PhysRevB.58.R14741,Greiner2002, 
 PhysRevLett.81.3108, L_hmann_2012, PhysRevB.79.174515, Bakr547, Weitenberg2011} {and the detection of anomalous transport in 2D quantum Heisenberg  magnets~\cite{PhysRevLett.113.147205}}
--- but has also faced unexpected obstacles in accessing the low-temperature regime of the  model in its original, fermionic, {\it avatar}. Another challenge is that much of the theoretical lore on the Hubbard model focuses on observables --- such as conductivities and spectral properties --- that are naturally accessed in solid-state experiments but are often less tractable from an atomic-physics perspective. Despite these hurdles, over the past few years  different groups have been able to access a range of temperature scales in Fermi-Hubbard optical lattices~\cite{Chiu251,Mazurenko2017}, and perfected new techniques, such as quantum gas microscopy,~\cite{Bakr2009,PhysRevLett.114.193001, Haller2015, PhysRevLett.114.213002, PhysRevLett.115.263001, PhysRevA.92.063406} that offer direct lattice-scale  probes of these systems.~\cite{Vijayan186,PhysRevLett.121.103001,Boll1257,Nichols383, doi:10.1146/annurev-conmatphys-070909-104059, PhysRevX.8.011053, PhysRevLett.122.153602,Brown1385}

In parallel, recent progress in the study of integrability applied to non-equilibrium systems has led to the formulation of `generalized hydrodynamics' (GHD).~\cite{PhysRevX.6.041065,PhysRevLett.117.207201,SciPostPhys.2.2.014,PhysRevLett.119.020602,PhysRevLett.119.220604, PhysRevB.97.045407, SciPostPhys.3.6.039, PhysRevLett.120.045301, PhysRevB.96.081118, PhysRevLett.120.164101, PhysRevB.100.035108, PhysRevLett.123.130602, 10.21468/SciPostPhys.8.3.041, PhysRevLett.124.140603, friedman2019diffusive, 2020arXiv200301702B, 2020arXiv200411030D, PhysRevB.101.035121, PhysRevX.10.011054, 2020arXiv200506242P} This is a systematic framework for treating the the effective long-wavelength fluctuations of integrable models, which is a convenient route to access their far-from-equilibrium transport and response properties.~\cite{2020arXiv200303334B,Vasseur_2016} As these are notoriously challenging to compute from first principles using Bethe ansatz techniques, GHD has dramatically simplified the application of tools from integrability to the computation of many experimentally-relevant observables. It has been applied, with notable success, to a variety of integrable systems such as the {Lieb-Liniger gas~\cite{PhysRevLett.120.045301, SciPostPhys.3.6.039, PhysRevLett.119.195301} or the XXZ spin-$\frac{1}{2}$ chain}.~\cite{PhysRevLett.117.207201,PhysRevB.96.115124, PhysRevLett.122.127202,PhysRevB.97.045407,Gopalakrishnan16250,Zotos2017,PhysRevB.97.081111} By building a Boltzmann-like kinetic theory for {stable} quasiparticles, GHD  has provided insights into the nature of transport and hydrodynamics in these systems. {Intuitively,  this kinetic approach remains valid even if the quasiparticle gas is not dilute, since scattering processes in integrable systems 
factorize.} {Recent developments} include explaining how diffusive corrections to ballistic quasiparticle motion arise microscopically,~\cite{PhysRevLett.121.160603,PhysRevB.98.220303, 10.21468/SciPostPhys.6.4.049, 2019arXiv191101995M} and identifying the physical origin of the universal superdiffusive dynamics observed numerically in systems with non-Abelian symmetries.~\cite{PhysRevLett.106.220601,Ljubotina:2017aa,PhysRevLett.121.230602, PhysRevLett.122.127202, PhysRevB.101.041411, PhysRevLett.123.186601,PhysRevLett.122.210602,  2020arXiv200106432D, 2020arXiv200313708D, 2019arXiv190905263A, PhysRevB.101.121106, 2019arXiv191201551D,2020arXiv200305957K}

Spurred by these developments, here we apply the techniques of GHD to the one-dimensional Hubbard model. We focus on two characteristic features of the one-dimensional model: (i) temperature-tuned spin-dynamics crossovers in the regime of ballistic transport at strong coupling; and (ii) superdiffusive dynamics at half filling and/or zero magnetization. In the former case, we identify an integrable analog of the crossover between spin-incoherent~\cite{RevModPhys.79.801,PhysRevLett.92.176401,PhysRevLett.92.106801, PhysRevB.72.125416,PhysRevB.73.165104,PhysRevB.95.024201, PhysRevB.75.205116} and spin-coherent dynamics identified within the framework of Luttinger liquid theory~\cite{Giamarchi2003}. We give a precise characterization of this crossover in the language of integrability, and identify its signature in the spin Drude weight (that characterizes the ballistic transport of spin). We also compute the Drude weights for charge and energy transport, in which the crossover is only manifest in subleading corrections in $1/U$. 
At half-filling and zero magnetization, some subset of conserved charges are transported sub-ballistically but super-diffusively, with dynamical properties governed by Kardar-Parisi-Zhang (KPZ) scaling~\cite{PhysRevLett.56.889}, while the energy transport remains ballistic. We present analytical, semiclassical, and numerical arguments for KPZ scaling at the special $SO(4)$ symmetric point, and complement this with a computation of the nonzero energy Drude weight.
 [Note that a previous study,~\cite{PhysRevB.96.081118} whose results we build on, has considered ballistic energy transport at half-filling and zero magnetization but did not discuss superdiffusion]. We thus give a comprehensive picture of temperature-dependent transport and response in the one-dimensional Hubbard model. The present discussion thus complements existing studies that have addressed transport in the one-dimensional Hubbard model  using rigorous  bounds on transport coefficients,~\cite{PhysRevLett.117.116401, PhysRevB.55.11029, PhysRevB.11.573, CARMELO2018418, CARMELO2013484} and via numerical simulations.~\cite{PhysRevLett.117.116401,Karrasch2017,PhysRevB.95.115148, PhysRevB.76.125110, PhysRevB.71.085110, Garst_2001, PhysRevB.70.205129, PhysRevB.90.155104, PhysRevB.92.205103, PhysRevB.86.125118} It also substantially extends previous GHD results~\cite{PhysRevB.96.081118, PhysRevLett.121.230602} by studying superdiffusion, crossovers in spin dynamics, and the associated experimental signatures. 
 {We emphasize that many of the distinctive experimental signatures of spin transport in the Hubbard model should be detectable in near-term experiments on ultracold atoms using optical gas microscopes.~\cite{Nichols383}}

The remainder of this paper is organized as follows. Two introductory sections provide background on the Hubbard model, its symmetries, and its exact solution in one dimension (Section ~\ref{sec:HubbardBasic} ) and  a summary of  techniques and results from GHD (Section~\ref{sec:GHD}). 
We have attempted to present a physically motivated introduction to these techniques; readers familiar with GHD and the Hubbard model can skip these sections.
Having laid the necessary groundwork, we then turn to an analysis of finite-temperature transport in the strong-coupling regime in Section~\ref{sec:nonhalffilling} before turning to superdiffusive transport at half-filling/zero magnetization in Section~\ref{sec:halffilling} Finally, we close with a summary of results and outlook for future work in Section~\ref{sec:conclusions}. We also include four technical appendices: Appendix~\ref{sec:app:TBA} provides more details on the TBA and is largely pedagogical, Appendix~\ref{sec:app:TBAsol} summarizes technical details of the solutions of the TBA equations, and Appendices~\ref{sec:app:largeM} and ~\ref{sec:app:largeU} summarize various asymptotic expansions used in the main text.

	\section{\label{sec:HubbardBasic}One-dimensional Hubbard model: overview and exact solution}
	\subsection{Model and Symmetries}
	Our focus throughout this paper will be the electronic Hubbard model, described by the Hamiltonian
\begin{equation}\label{eq:HubHam}
H = \hat{T} + \hat{V}  - \mu \hat{Q} -h\hat{S}^z,
\end{equation}
where
\begin{equation}\label{eq:hop}
\hat{T} =  -t\sum_{j,\sigma=\uparrow,\downarrow} \left(c^\dag_{j+1,\sigma} c_{j,\sigma} + \text{h.c.}\right),
\end{equation}
is a nearest-neighbor hopping term (we set $t=1$ henceforth),
\begin{equation}\label{eq:int}
\hat{V} = U \sum_j\left(n_{j,\uparrow}-\frac{1}{2}\right)\left(n_{j,\downarrow}-\frac{1}{2}\right)
\end{equation}
is the usual on-site Hubbard interaction (with $n_{j,\sigma} \equiv c^\dagger_{j,\sigma} c_{j,\sigma}$), and the chemical potential $\mu$ and magnetization $h$ couple to the two $U(1)$ conserved quantities, namely the total electric charge
\begin{equation}
\hat{Q} =  \sum_j \left( n_{j,\uparrow} + n_{j,\downarrow}\right), 
\end{equation}\label{eq:Qdef}
and  total magnetization along the $z$-axis
\begin{equation}\label{eq:Szdef}
\hat{S}^z =  \frac{1}{2}\sum_j \left(n_{j,\uparrow} - n_{j,\downarrow}\right),
\end{equation}
whose transport, along with that of the energy, will be our primary concern below. We have chosen a convention such that  for $\mu=0$ the system is at half-filling, which for a chain of $L$ sites is defined as $\frac{\langle \hat{Q}\rangle}{L}  =\frac{1}{L} \sum_{j,\sigma} \langle n_{j,\sigma} \rangle = 1$. 

Besides its evident translational invariance, the Hubbard Hamiltonian $H$ \eqref{eq:HubHam} enjoys several global symmetries; for a complete treatment  we refer the reader to Ref.~\onlinecite{essler_2005} and only summarize those most pertinent to our discussion. First, observe that $H$ commutes with $\hat{Q}$ and  $\hat{S}^z$ for {\it all} values of $h$ and $\mu$, and so these are always symmetries: below, we will discuss the transport of the conserved charge and magnetization corresponding to these two $U(1)$ symmetries. However, the global symmetry  is enhanced when either $\mu=0$ or $h=0$ (or both). 
 For $h=0$, the $U(1)_s$ spin symmetry of rotations about the $z$ axis extends to a full non-abelian $SU(2)_s$ symmetry of rotations about an arbitrary axis in spin space. This $SU(2)_s$ symmetry can can be made manifest~\cite{fradkin_2013} by rewriting the interaction term as $\hat{V} = -\frac{2U}{3} \left(\bs_j \cdot \bs_j \right)$, where we have defined $\bs_j = \sum_{\alpha,\beta} c_{j\alpha}^\dag \frac{\boldsymbol{\sigma}_{\alpha\beta}}{2}c_{j\beta}$, where $\boldsymbol{\sigma} = (\sigma_x,\sigma_y,\sigma_z)$ is a triplet of Pauli matrices. {Evidently, $\hat{S}^z$ coincides with our definition in \eqref{eq:Szdef}, and the other components of $\boldsymbol{S}$ are chosen so as to satisfy the usual $SU(2)_s$ Lie algebra $[\hat{S}^\alpha, \hat{S}^\beta] = i\epsilon^{\alpha\beta\gamma} \hat{S}^\gamma$ of spin rotations.} As a consequence of this $SU(2)_s$ symmetry, thermal states for $h=0$ are not magnetized in any direction.
On the other hand, for $\mu=0$, the nearest-neighbor model has a distinct $SU(2)$ invariance discovered by Yang~\cite{PhysRevLett.63.2144}
 and dubbed the `$\eta$-pairing' symmetry.~\footnote{The $\eta$-pairing symmetry is most conveniently understood by performing a  particle-hole (or `Shiba') transformation on a single spin species. This interchanges charge and spin, and maps the Hamiltonian $H$ (upto unimportant constants) to another Hamiltonian $H'$ of the same form as \eqref{eq:HubHam} but with  $U'=-U$, $h'= -\mu$ and $\mu'=-h$. The interchange of $SU(2)_\eta$ and $SU(2)_s$ generators  reveals that $SU(2)_\eta$ invariance of $H$ is equivalent to an $SU(2)_s$ invariance of its single-spin-Shiba-transform $H'$.} The three generators  $\hat{\be} = (\hat{\eta}^x, \hat{\eta}^y, \hat{\eta}^z$) of the $SU(2)_\eta$ symmetry take the form
\begin{equation}
\hat{\eta}^x= \frac{\hat{\eta}^{+}\!+\!\hat{\eta}^{-}}{2}, \hat{\eta}^y= \frac{\hat{\eta}^{+}\! -\!\hat{\eta}^{-}}{2i}, \hat{\eta}^{z} = \sum_j \!\frac{n_{j\uparrow} \!+ \!n_{j\downarrow} \!-1}{2},
\end{equation}
where $\hat{\eta}^+ = -\sum_j (-1)^j c_{j\uparrow}^\dag c_{j\downarrow}^\dag = (\hat{\eta}^{-})^\dag$. It is straightforward to show that this generates an $SU(2)_\eta$ algebra $[\hat{\eta}^\alpha, \hat{\eta}^\beta] = i\epsilon^{\alpha\beta\gamma} \hat{\eta}^\gamma$ that is distinct from that of spin rotations, since $[\hat{S}^\alpha, \hat{\eta}^\beta] =0$. From the relation $\hat{Q} = 2\hat{\eta}^z + 1$, it is clear that for $\mu\neq0$, the Hamiltonian only has the $U(1)$ symmetry generated by $\hat{\eta}^z$, which coincides with that of charge conservation.
However, when $\mu=0$, the system enjoys the full $SU(2)_\eta$ symmetry generated by the above operators. Therefore as in the case when $h=0$, the extra $SU(2)_\eta$ symmetry has implications for the thermal states, as $\langle \eta^z\rangle=0$, thermal states for $\mu=0$ are at half-filling.

Finally, at the special point  $\mu=h=0$ which lies at the intersection of the lines of $SU(2)_s$ and $SU(2)_\eta$ symmetry, the Hubbard Hamiltonian enjoys an extended $SO(4)\simeq SU(2)_s\times SU(2)_\eta/\mathbb{Z}_2$ symmetry.~\footnote{\label{fn:Z2explanation}The $\mathbb{Z}_2$ quotient reflects the fact that although $[\hat{S}^\alpha, \hat{\eta}^\beta] =0$, the allowed irreducible representations of $SU(2)_s$ and $SU(2)_\eta$ are not independent:  either both are integer or both are half-odd-integer. Note that this distinction is only important in considering the global Lie group structure rather than the Lie algebra, and is hence unimportant to our semiclassical analysis in Section~\ref{sec:softgauge}.}

Note that the symmetries discussed up to this point are not necessarily specific to the nearest-neighbor Hubbard model or to its one-dimensional setting. Absent an explicit breaking of spin rotation (e.g. by the introduction of spin-orbit coupling), even extended Hubbard models continue to enjoy $U(1)_s$ ($SU(2)_s$) symmetry for $h\neq0$ ($h=0$). Similarly, the global $U(1)_c$ symmetry is generically a feature of Hubbard-like models, unless an explicit superconducting pairing term is introduced, for instance in order to capture the effects of externally induced superconductivity. Finally, for any {\it bipartite}~\footnote{By bipartite hopping we mean that the lattice may be divided into two disjoint sets of sites $A, B$ such that $\hat{T}$ has no matrix elements between the two sets, which automatically forces $\mu=0$.} hopping $\hat{T}$ we expect the full $SU(2)_\eta$ symmetry.

However, as noted in the Introduction, the one-dimensional nearest-neighbor Hubbard model -- unlike its generalizations and higher-dimensional counterparts ---  is an integrable model that hosts an extensive set of conserved quantities.  Consequently  we may determine its full spectrum of eigenstates  {\it exactly} for any fixed system size $L$,  particle number $N$, and magnetization $M$ via the (nested) Bethe ansatz. By taking the thermodynamic limit of the resulting Bethe equations and using the framework of generalized hydrodynamics, we can extract {transport coefficients} such as Drude weights and {d.c. conductivities}.  Henceforth, we focus on the one-dimensional model; in the remainder of this section we briefly summarize the nested Bethe ansatz and its thermodynamic limit.

	\subsection{\label{sec:TBA}Thermodynamic Bethe Ansatz}
	
	The key idea of the Bethe ansatz is to construct eigenstates  in the occupation-number representation of one or more species of quasiparticle excitations above a reference vacuum state (for example, the state with all spins down in a Heisenberg spin chain). Each quasiparticle excitation can be ascribed  a pair of labels that respectively describe its species and its {quasimomentum}, both of which are preserved in collisions~\footnote{Heuristically, when a pair of  quasiparticles collide,  they `pass through' each other picking up a scattering phase shift that depends on the species indices of both quasiparticles and on their rapidity difference.}. The latter is not precisely the physical momentum of the excitation (the distinction is explained below), but  plays a role similar to the momentum in organizing the spectrum.  It is frequently useful to reparametrize the quasi-momentum in terms of a complex-valued quantity known as the rapidity. 	

	 The essence of integrability is that all multi-particle scattering processes can be factorized into combinations of two-particle scattering events; this in turn is a consequence of the existence of an infinite  number of local conserved charges. Translation invariance, the phase shifts due to quasiparticle scattering,  and the boundary conditions combine to constrain the allowed rapidities. The relevant constraints are encoded by  set of algebraic `Bethe equations' satisfied by the admissible rapidities, termed `Bethe roots'.  Rapidities (or equivalently, quasi-momenta) play a role similar to momenta in free-particle systems; however, a crucial difference is that the allowed values of rapidity (quasi-momenta)  of any given quasiparticle is influenced by the presence of all the other particles in the system. It is this nontrivial feedback that is captured by the Bethe equations.
	 	
Except for the simplest Bethe-ansatz solvable models (such as the Lieb-Liniger gas with repulsive interactions),  the Bethe roots are generically complex. However, a simplification is afforded by the so-called {\it string hypothesis}: namely, that in rapidity space the Bethe roots cluster into `strings'  that share the same real part, and correspond in real space to a set of bound states of quasiparticles. This hypothesis is approximate for finite systems but is believed to become exact in the thermodynamic limit ($N,L\to \infty$ with $N/L$ fixed). In this limit, the structure of roots admits the following simple interpretation: strings are bound states of elementary quasiparticles, and propagate as stable composite entities with a well-defined dispersion relation. In this section, in order to orient the discussion in the rest of the paper, we briefly summarize the key physical features of the thermodynamic Bethe ansatz solution of the one-dimensional Hubbard model. A more extensive discussion is in Appendix~\ref{sec:app:TBA}.
		
	First, since $U(1)$ charge and spin conservation are valid  symmetries for any $\mu, h$, we can work in sectors with fixed particle number  $N = N_\uparrow + N_\downarrow$ and magnetization $M =\frac{N_\uparrow-N_\downarrow}{2}$,  where $N_\uparrow$, $N_\downarrow$ are the number of up and down spin electrons respectively. Then, exploiting particle-hole symmetry $\mathcal{P}: c_{j,\sigma} \mapsto (-1)^j c^\dagger_{j,\sigma}$ (under which $\mu\mapsto-\mu$), we can restrict ourselves to sectors with the total number of particles $N$ satisfying $N<L$.
	Similarly, exploiting the discrete symmetry $S^z\mapsto-S^z$ (under which $h\mapsto -h$), we can limit our study to sectors for which the magnetization $M>0$, Under these assumptions, we can build a basis of Bethe ansatz states by starting with states of $N$ spin-up electrons, whose rapidity we denote by $u_j$ (where $1\leq j\leq N$), and adding $N_\downarrow$ magnon-like excitations, with rapidities $w_j$ (where $1\leq j\leq N_\downarrow$). 
	 We can also associate each root with a definite charge  under the two $U(1)$  symmetries: each $u_j$ root has charge $q=1$ and $z$-magnetization $m=1/2$, while  each $w_j$ root has charge $q=0$ and $z$-magnetization $m=-1$. Note that there is formally a slight subtlety with the Bethe ansatz states constructed in this manner: they correspond to only the `highest weight' states in each $SU(2)_s$, $SU(2)_\eta$ sector (as defined in the $h=\mu=0$ limit). In each sector the remaining states in the spectrum must be generated by acting on the Bethe-ansatz states with $\hat{S}^- = \hat{S}^x-i \hat{S}^y$ and $\hat{\eta}^-$. However, as we explain in Appendix \ref{sec:app:TBA} this is unimportant in the thermodynamic limit as the `missed'  states only contribute a logarithmically vanishing correction to the  free energy density.

Assuming the string  hypothesis, the Bethe ansatz  spectrum of the Hubbard model is built of an infinite number of quasiparticles/strings species that can be broadly classified into one of three types:
	\begin{description}
		\item[$y$-particles] Spin-up electrons not bound into larger objects. $q_y=1$ and $m_y=1/2$; these come in two branches,~\footnote{Note that the fact that there are two branches is because we have chosen to label particles by rapidities; quasimomenta are multivalued functions of rapidity, and so we need an extra label to keep track of the relevant quasi-momentum branch when working with rapidities.} labeled $\pm$
		\item[$M|w$-strings] with $M\in\mathbb{N}$, $M\geq1$. Strings of $M$ $w$-roots, corresponding to a magnon of length $M$. $q_{M|uw}=0$ and $m_{M|uw}=-M$.
		\item[$M|uw$-strings] with $M\in\mathbb{N}$, $M\geq1$. Strings of $2M$ $u$-roots and $M$ $w$-roots, forming a spin-singlet object. $q_{M|uw}=2M$ and $m_{M|uw}=0$.
	\end{description}
 We will refer to these three objects in the TBA spectrum as `$y$-particles', {`magnons' and  `singlets' }, respectively. Note that there is an infinite number of magnon and singlet species, indexed by positive integers.
	
	As noted above, each quasiparticle/string is labeled by its species and by a rapidity that describes the position of the corresponding Bethe root. The advantage of working with strings rather than individual Bethe roots is that string centres (which we denote by $u$) are real, and hence easier to handle than the full set of complex Bethe roots.  [We will  use  `quasiparticle' and `string' interchangeably, but the meaning will be clear from the context.]  
	
 For a large number of particles,  the  Bethe equations rapidly become intractable. Fortunately, in the thermodynamic limit (taken in the sense of $N, L\to \infty$ with $N/L$ held fixed) it is unnecessary to keep track of the position of individual Bethe roots. 
 Instead,  it is convenient to switch to a description in terms of their densities in rapidity space. These are conveniently captured by appropriate rapidity-space quasiparticle distribution functions.
This description, that combines the simplifications afforded by statistical mechanics with the exact results of the Bethe ansatz   is known as the thermodynamic Bethe ansatz~\cite{takahashi_1999,essler_2005,PhysRevB.96.081118} (TBA). The basic idea behind the TBA is to construct a thermal `generalized' Gibbs state for an integrable model by applying the maximum entropy principle, but constrained on holding fixed the values of an extensive set of conserved quantities. The latter explains why this Gibbs ensemble is `generalized' --- it involves an extensive set of Lagrange multipliers, one for each conserved quantity. 

{
A generalized equilibrium state can be consistently defined in terms of a vector of  generalized `filling factors' for quasiparticles of different species and rapidities $\bn = \{n_a(u)\}$, where
	\begin{equation}
	n_a (u)  = \frac{1}{1+Y_a(u)}.
	\end{equation}
	 The set of functions $\{Y_a(u)\}$ completely characterize the state, with $1/Y_a(u)$ analogous to a Boltzmann factor for the quasiparticles. A state of a given species at an allowed rapidity  can either be occupied by a quasiparticle or empty (`occupied by a hole'), explaining the formal resemblance of the filling factor to a fermionic occupation probability. We introduce the total density of quasiparticle state $\rho_a^t (u)$, in terms of which the density of {\it occupied} quasiparticle states (usually termed the particle density) is given by $\rho_a(u) = n_a (u) \rho^t_a(u)$. Frequently, a complementary quantity termed the {\it hole} density  $\bar{\rho}_a =\rho^t_a -\rho_a$ is also defined, as well as a corresponding hole filling factor, $\bar{n}_a(u) = \bar{\rho}_a(u)/ \rho^t_a(u) = 1-n_a(u)$.

 
 \section{\label{sec:GHD}Generalized Hydrodynamics}
		
The TBA framework outlined above allows one to characterize equilibrium states of integrable systems, but does not offer direct access to correlation functions, transport, or other dynamical properties. To treat such questions exactly, one is forced to use form-factor expansions that are generally intractable. However, the framework of GHD offers a way to leverage the relatively simple TBA solutions to predict the coarse-grained dynamics of integrable systems. We now quickly sketch this framework; for a more detailed account see Ref.~\onlinecite{doyon2019lecture}. 

GHD is built on the assumption that the system can be partitioned into mesoscale regions of size $l$, each of which is approximately in a local equilibrium state (i.e., one described by TBA); globally, the system is away from equilibrium because the chemical potentials vary from cell to cell. Under this hydrodynamic assumption, the coarse-grained dynamics of the system reduces to the dynamics of the parameters that specify a local TBA state, for example its quasiparticle densities. The quasiparticle densities evolve according to two sets of generalized hydrodynamic  equations: (i)~a continuity equation for quasiparticle densities of each species and rapidity, $\partial_t \rho(\lambda, x, t) + \partial_x j(\lambda, x, t) = 0$; and (ii)~a constitutive relation, which posits that each quasiparticle moves ballistically with its effective group velocity $v^\mathrm{eff} [ \rho]$. This constitutive relation reads: $j(\lambda, x, t) \equiv \rho(\lambda,x,t) v^{\mathrm{eff}}[\rho](\lambda,x,t)$. After some algebra, these hydrodynamic equations can be rewritten in the following advective form, in terms of the filling factors $n(\lambda)$~\cite{PhysRevX.6.041065,PhysRevLett.117.207201}:
\begin{align}\label{eq:GHDadvection}
		\partial_t n_a(u) + v^{\text{eff}}_a[\bn(x,t)](u) \partial_x n_a(u)=0.
\end{align} 
{
The GHD equation~\eqref{eq:GHDadvection} captures the evolution of quasiparticle densities as one goes from local to global equilibrium states. In general, this evolution is nonlinear, as $v^{\text{eff}}$ for each quasiparticle depends on the occupation numbers of all the others. More precisely, we have $v^{\text{eff}}_a \equiv (e'_a)^{\mathrm{dr}}/(k_a')^{\mathrm{dr}}$, where $e_a$ and $k_a$ are the bare energy and momentum of the string $a$, respectively; and $(\dots)^{\mathrm{dr}}$ refers to a dressing operation of these quantities in a given (generalized) equilibrium state $\bn$, described more quantitatively in Appendix~\ref{sec:app:TBA}. In this paper we  restrict ourselves to linear response, for which it suffices to consider small fluctuations about a spatially homogeneous generalized Gibbs state.}

Eq.~\eqref{eq:GHDadvection} gives a prescription for computing the dynamics of the local occupation factors; the remaining step is to relate these back to physical observables. To do so we must reconstruct the local TBA state, given all the occupation numbers. As a simple example, consider the equilibrium correlation functions of local charge densities, $\langle q_i(x,t) q_j(0,0) \rangle - \langle q_i \rangle \langle q_j \rangle$ where $i, j$ index the infinitely many conserved charges. This correlation function is proportional (via the fluctuation-dissipation theorem) to the charge response at $(x,t)$ due to a slight change in the chemical potential $\mu_j$ for charge $j$ in the hydrodynamic cell at $(0,0)$. 
{Quasiparticle $a$ with rapidity $u$ carries a dressed charge $(q^i_a)^{\rm dr}(u)$ for the $i$th conserved quantity, which is dressed by interactions in a given background GGE.} 

{The physical picture that emerges from these equations is simple: each quasiparticle carries dressed charge $(q^j)^{\mathrm{dr}}$, and propagates at velocity $v^{\mathrm{eff}}$. Thus the connected component of dynamical correlation functions for charge obey the equation~\cite{SciPostPhys.3.6.039,PhysRevB.96.081118}
\begin{align}
\langle q_i(x,t) q_j(0,0) \rangle_{{c}} & = \sum_a \int du \rho_a(u) (1 - n_a(u)) \times \nonumber \\ &  (q^i_a)^{\mathrm{dr}}(u)(q^j_a)^{\mathrm{dr}}(u) \delta[x -   v^\mathrm{eff}_a(u) t].
\end{align}
The correlation functions for a generic operator can be inferred from this result by the following reasoning: in the hydrodynamic limit, all correlation functions other than those of conserved charge densities decay rapidly. Therefore, to find the correlation functions of the operator, one simply needs to compute its overlap with each conserved charge, and then apply the previous result. }

{A quantity of particular importance is the Drude weight, defined as the long-time limit of the current-current correlation function, $\langle J_i(t) J_k(0)\rangle$. The current operator can be written as $J_i = A_{ij} Q_j + \ldots$, where $\ldots$ represents the projection of the current onto fast operators, and the matrix $A_{ij}$ can be related to the dressing transformation and the effective velocity in hydrodynamics. The Drude weight, then, is $D = A_{ij} \langle Q_i Q_j \rangle A_{jk}$. Once again, by expressing these matrix products in the basis of $n(\lambda)$, one arrives at the result~\cite{SciPostPhys.3.6.039,PhysRevB.96.081118}:
	\begin{equation}
		D_{i j} =\label{eq:generalDrude} 
		\beta\sum_{a} \int du \, \rho_a  
		\bar{n}_a (q^i_a)^{\text{dr}} (q^j_a)^{\text{dr}}  \left[v^{\text{eff}}_a\right]^2,
	\end{equation}
providing a closed-form expression for the Drude weight solely from TBA data. Once again this expression has a rather simple physical interpretation: a quasiparticle of type $(a, u)$ carries charge  $(q^j)^{\mathrm{dr}}$ while moving ballistically at a speed $v^{\mathrm{eff}}$. Since the quasiparticle never scatters, this current does not relax. The Drude weight is the sum of these persistent currents due to each quasiparticle. }

{We will be interested here in the Drude weights and correlation functions of energy, charge, and spin in the Hubbard model. We adopt the standard terminology where the diagonal terms $i=j$ are referred to as {\it the} conductivity/Drude weight of conserved charge  $\mathcal{O}$ (and use a single label), whereas for $i \neq j$ they are called the cross-conductivity/crossed Drude weight.} We focus primarily on the former, although we briefly discuss crossed Drude weights in the spin-incoherent Luttinger liquid regime. 
	
We note that there is a choice of convention in computing the energy Drude weight. We can either compute the Drude weight corresponding to the full Hamiltonian $H$ (bare energy $={e_a}(u)$), or  to the `reduced' Hamiltonian without chemical potential/magnetization terms $\tilde{H} = \hat{T} +\hat{V}$,  (bare energy $\tilde{e}_a(u)$), where the choice of bare charge then carries over to the dressed charges. Since the relation between $\tilde{e}_a$ and $e_a$ is comprised of conserved charges, this means that the Drude weight computed for $e_a$ will involve contributions from the spin, charge, and all the crossed weights due to the $[e_a^{\text{dr}}(u)]^2$ term in \eqref{eq:generalDrude}. Therefore to simplify matters we compute the `reduced' energy Drude weight corresponding to $\tilde{H}$. [In order to convert this to the full energy Drude weight of $H$ we must also compute the crossed Drude weights using the methods presented here.]

To avoid confusion, henceforth we denote by $\mathcal{O}_a$ the conserved charge carried by  quasiparticles of species  $a$, and focus on the electric charge, the magnetization, and the (reduced) energy, viz. $\mathcal{O} =  q, m, \tilde{e}$.

\section{\label{sec:nonhalffilling}Ballistic transport, Drude weights, and spin-coherence crossovers at strong coupling}

We are now ready to address one of our two main objectives: to analyze the structure of transport processes in the Hubbard model in the strong coupling regime $U/t\gg1$. As noted in the Introduction, the lines $h=0$, $\mu=0$ require special consideration due to the presence of non-Abelian symmetries, which lead to a transport regime that is intermediate between ballistic transport with nonzero Drude weight, and simple diffusion. Accordingly, we discuss this regime in the next section and for now focus on the case when $\mu\neq 0$ and $h\neq 0$.

In the strong coupling limit, a hierarchy of well-separated energy scales can be identified, allowing us to distinguish four different regimes (Fig.~\ref{fig:regimes}) depending on the temperature $T$. [Note that since we have fixed $\mu, h\neq0$, within each regime we must be careful to compare the temperature scale with those set by the chemical potential and the field; we provide further details on this below.]

Starting from high temperature,  the first transport regime we encounter is
\begin{enumerate}[label=(\roman*)]
\item $\boldsymbol{T\gtrsim U \gg t}$: this corresponds to `generic' high temperature transport, to which all string types contribute. 
Systems at weak- and strong-coupling  show qualitatively similar behavior in this limit.
\end{enumerate}
We access the remaining regimes by lowering the temperature so that $U\gg T$. Transport in these regimes can be approximately understood by projecting out double occupancies to obtain an effective $t-J$ model~\cite{essler_2005} with $J\sim t^2/U\ll t$, which we can subdivide further into three regimes:
\begin{enumerate}[resume,label=(\roman*)]
 \item $\boldsymbol{U\gg T \gg t \gg J}$: in this case, away from half filling we have $\mu=-{O}(U)$, so that $uw$-strings are not thermally occupied and drop out of transport; therefore, we expect transport properties to be comparable to that of the $t-J$ model at $T=\infty$. 
 \item  $\boldsymbol{U\gg t\gg T \gg J}$:  this ordering of scales leads to an unusual situation in which charge degrees of freedom are  in the low-temperature phase (effectively at $T\simeq0$), whereas the spin degrees of freedom remain high-temperature (i.e. approximately at $T=\infty$.) A similar regime has been identified in the context of generic Luttinger liquid theory (i.e., without any assumption of integrability) where it has been dubbed the spin-incoherent Luttinger liquid (SILL).~\cite{RevModPhys.79.801}
 \item  $\boldsymbol{U\gg t\gg J\gg T}$:  when $T$ is the lowest  scale in the problem, we expect to recover  normal Luttinger liquid-like behavior including the identification of two distinct speeds that control ballistic propagation of spin and charge, as in simpler integrable models~\cite{PhysRevB.99.014305} {(see also Ref.~\onlinecite{PhysRevLett.120.176801,Bertini_2018})}. As in  the case of regime (i), we do not expect a qualitative distinction between weak and strong coupling in this low-temperature regime.
\end{enumerate}

		\begin{figure}
		\includegraphics[width=0.95\linewidth]{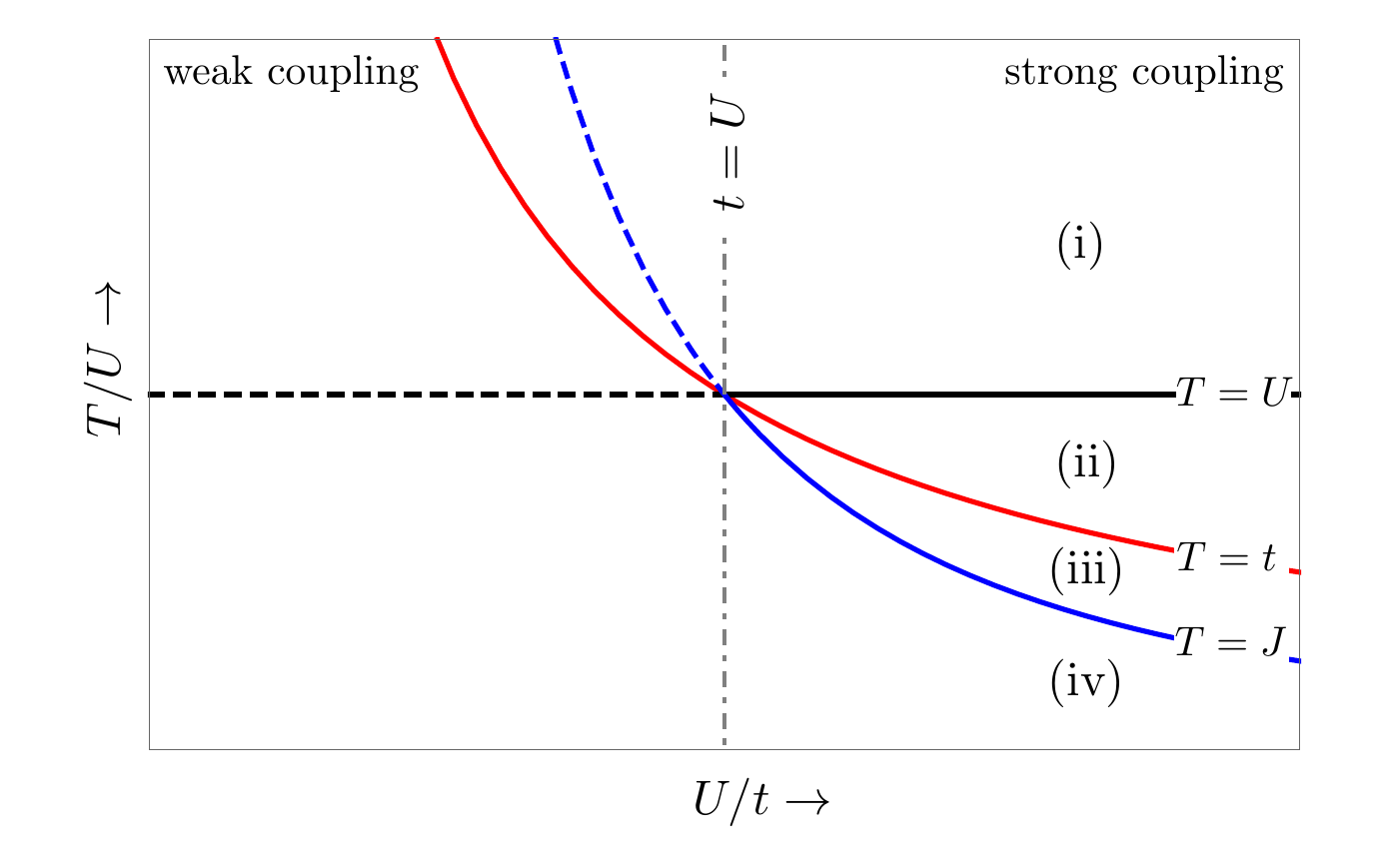}
		\caption{\label{fig:regimes} Regimes of transport for the Hubbard model. At strong coupling $U\gg t$ we can distinguish four temperature regimes delineated by sharp crossovers (indicated by the solid lines) in dynamics. In descending order of temperature$T$ these are (i) the `high temperature Hubbard' regime, where $T$ is the biggest energy scale; (ii) the high-temperature $t-J$ regime, where we can effectively ignore double-occupancies since $U\gg t$, but $T$ still exceeds both the charge scale $t$ (i.e., the holon bandwidth)  and the effective spin-exchange scale $T\gg J\sim t^2/U$; (iii) the `spin-incoherent' regime, where the charge fluctuations of the $t-J$ model are cold ($T\ll t$) but the spins remain hot, ($T\gg J$); and finally, the low-temperature  regime where the system is described as spin-charge separated Luttinger liquid of coherent charge and spin degrees of freedom, where $T$ is the lowest energy scale. At weak coupling, regimes (i) and (iv) are broadly similar and we expect a crossover at $T\sim t$. However, the weak-coupling crossovers for $t\sim J$ and $t\sim U$ are less significant  and hence we do not discuss them further in this work.}
	\end{figure}

For completeness, we briefly comment on the physics at weak coupling, $t\gg U$. First, as noted above behavior in the regimes (i) (which now emerges when $T\gg t$ and is again the largest energy scale) and (iv)  (where $T$ is the smallest scale in the problem) are broadly similar to that seen strong-coupling limit. There is no analog of the `high-temperature $t-J$ model' regime (ii), and the spin-incoherent regime (iii) is also absent in the sense discussed above.\footnote{{However, there is an intermediate regime in which $U \ll T \ll t$. In this regime spin excitations are effectively at high temperature, since their characteristic energy scale at weak coupling is the exchange scale $U$. We expect this regime to have some thermodynamic and transport features in common with the SILL.}}
We do not address this regime further in the present work (but see Ref.~\onlinecite{essler_2005})

		\begin{figure*}
		\includegraphics[width=\linewidth]{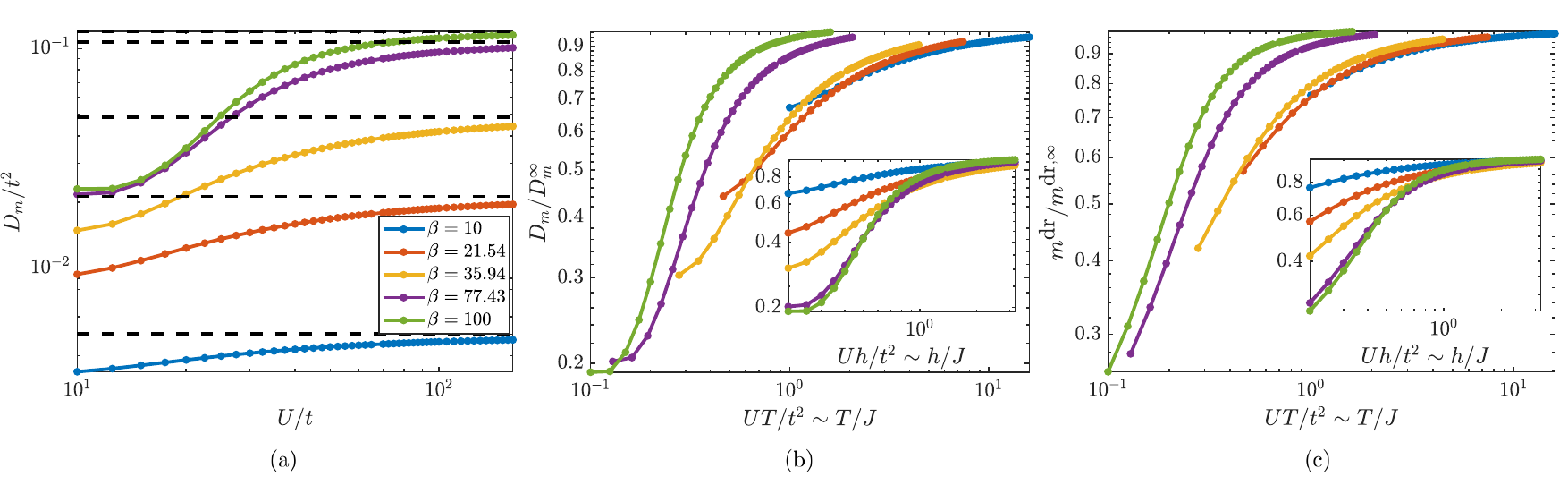}
		\caption{		\label{fig:spin-spin-drude}
			Spin Drude weight $D_m$ at the thermal crossover between spin-incoherent and spin-coherent  regimes  ((iii) and (iv) respectively in Fig.~\ref{fig:regimes}). We fix $\langle \hat{Q} \rangle/L=0.3$ and $h/t=0.04$.
			(a) $D_m$ as a function of $U$ for various $\beta$. Dashed line indicates the asymptotic value $D^\infty_m$ for $U\to\infty$ in regime (iii).
			(b) The same data as in panel (a). We observe that for $\beta h \lesssim 1$, $D_m/D^\infty_m$ departs from $1$ when $J\sim T$. Instead, if $\beta h \gtrsim 1$, $D_m/D^\infty_m$ departs from $1$ when $J\sim h$ (inset).
			{(c) We highlight that the crossover in the Drude weight is a consequence of a change of the dressed magnetization of the $y$-particles at the Fermi points.}
		}
	\end{figure*}

	\begin{figure*}
		\includegraphics[width=\linewidth]{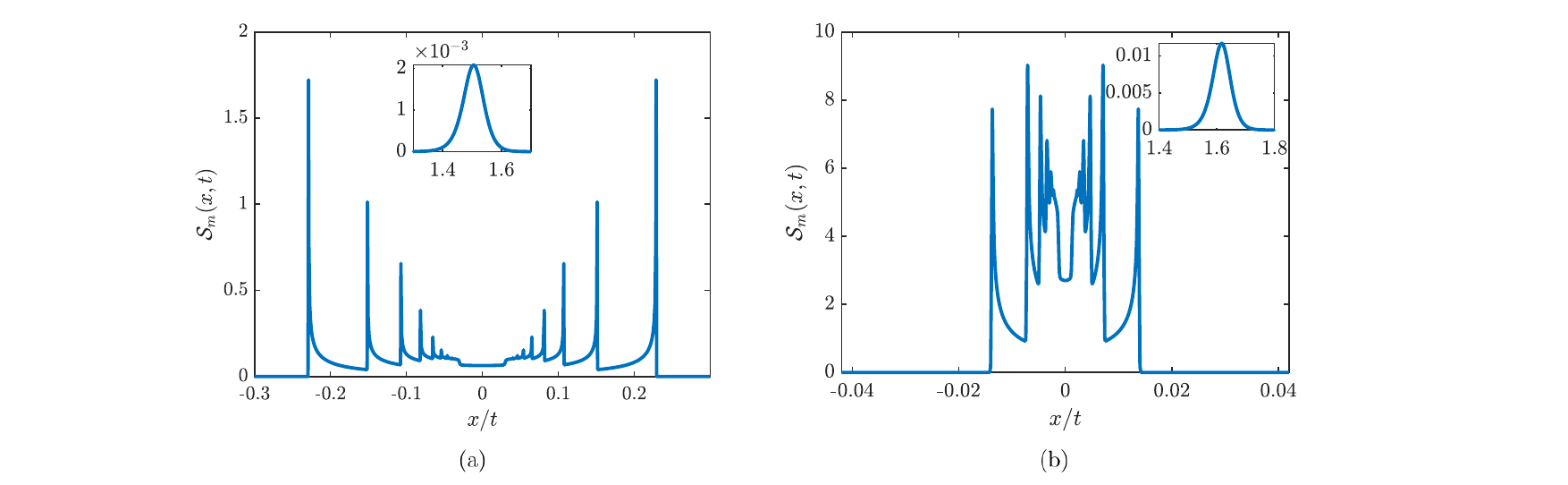}
		\caption{\label{fig:spin-spin-corr}Dynamical spin-spin correlation function in (a) the spin-coherent regime ($U=10$) and (b) the spin-incoherent regime ($U=160$). In each panel, the main figure displays the rich structure due to slow magnon modes, while the insets show the fast-moving leading front due to $y$-particles. \revisionchange{The peaks corresponding to $1|w$ and $2|w$ magnons are marked in the figure.} In both figures $h=0.04$, $\beta=35.9$, and $\langle \hat{Q} \rangle/L=0.3$.}
		\label{fig:ss-corr-U}
	\end{figure*}

	\subsection{Spin transport and spin-coherence crossovers}
Away from the lines $\mu=0, h=0$, the density is fixed away from half-filling  $\langle\hat{Q}\rangle/L \neq 1$ and the magnetization is fixed and non-zero $\langle \hat{S}_z\rangle/L \neq 0$.  
	Charge and energy transport are completely unaffected by the crossover from (iii) to (iv), in agreement with a general conjecture,~\cite{PhysRevB.72.125416} and confirmed by explicit calculation below. However,  spin transport is sensitive to the crossover, as we now show. In a non-integrable model the crossover would be very clear-cut, as in regime (iii) spin transport would not be ballistic, while in the Luttinger liquid regime (iv) we expect a non-zero spin Drude weight.  What happens at the crossover in the integrable Hubbard model is less obvious, since we expect the spin transport to have a ballistic (Drude) response at all temperatures. Surprisingly, the spin-incoherent to spin-coherent crossover has a sharp signature in the spin Drude response itself, as we now demonstrate.
	
 In the spin-incoherent regime (iii), the leading contribution in $t/U$ to the spin Drude weight comes from the $y$-particles and is given by the expression~\eqref{eq:large-U-low-T-Drude} with $\mathcal{O}_F$ replaced by the dressed magnetization at the Fermi points $m^{\text{dr}}_F=\tanh\left(\beta h/2\right)/2$. Regime (iv) is difficult to understand analytically due to the presence of a non-zero field, especially if $\beta h \lesssim 1$. Thus, we first analyze which parameters can affect the crossover and then rely on numerical solution of the TBA equation in cases where there can be a non-trivial crossover.
	
	The  parameters relevant to the characterization of spin transport are $\beta h$ and $h/J$, with the crossover (iii)-(iv) taking place at $\beta J \sim 1$. A first consequence of this observation is that the crossover (iii)-(iv) is more naturally observed by varying $U$ at a fixed $\beta$, since otherwise spin transport will already have a non-trivial dependence due to the variation of $\beta h$. We first analyze the case where $\beta h\gtrsim 1$. In this situation, regime (iii) is practically spin-coherent since the external field $h$ dominates the exchange scale $J$. As a consequence, we do not expect to see a sharp signature in spin transport at the (iii)-(iv) crossover, since the exchange scale is no longer relevant to the spin physics. Instead, we expect a crossover when, as $U$ decreases, $J$ becomes comparable with $h$ -- which occurs inside regime (iv) (see inset of Fig.~\ref{fig:spin-spin-drude}b). 
	However, we do expect non-trivial behavior at the (iii)-(iv) crossover when  $\beta h\lesssim 1$. In Fig.~\ref{fig:spin-spin-drude}, we demonstrate that around this parameter regime a crossover is indeed observable in the spin-Drude weight, {as} determined by numerically solving the TBA equations {(see Appendix~\ref{sec:app:TBAsol})}.
	
		In order to shed further light on this crossover, it is useful to examine its qualitative features in the dynamical spin-spin correlators $\mathcal{S}_m(x,t)$, shown in Fig.~\ref{fig:spin-spin-corr}. In both regimes, as noted above, the current response is dominated by fast $y$-particles, which produce a peak at $x/t\simeq v_F$ (see insets). 
		However, the spin-spin correlators also present a rich structure at smaller $x/t$, which is produced by the slow magnons.
		First, in the spin-incoherent regime (iii) a hierarchy of magnons (truncated at a length $M\sim T/h$) produces a structure which is overall peaked at small $x/t$ (see Fig.~\ref{fig:spin-spin-corr}b): in other words, the longest and slowest magnons (with $M\sim T/h$) give the dominant contributions to $\mathcal{S}_m$. In this regime we  expect to observe similar phenomenology  to that discussed in Ref.~\onlinecite{Gopalakrishnan16250} for the Heisenberg XXX chain. In contrast, in the spin coherent regime (iv), the amplitudes of the peaks due to $M|w$-magnons with $M>1$ tend to $0$ as $T$ decreases. This happens irrespective of the field $h$: if $h/T\gtrsim1$, $n_{M|w}\simeq 0$ for $M>2$, otherwise, if $h/T\lesssim1$, $\rho^t_{M|w}\to0$ as $T$ is lowered.~\cite{essler_2005,PhysRevB.9.2150} Thus, deep in regime (iv), $\mathcal{S}_m$ is dominated by $y$-particles and $1|w$-magnons alone, as can already be seen for the parameters in Fig.~\ref{fig:spin-spin-corr}a.
		However, as these results are most clearly manifest in the long-time limit (recall that the magnons are slow!) they might not be easy to observe in real-time dynamics on shorter timescales. Note that the change in the magnon properties across the crossover is not {\it directly} visible in the spin Drude weight, which is dominated by $y$-particles in both regimes (iii) and (iv). Instead, they affect the spin Drude weight indirectly, via the the change in the nature of the dressing of the $y$ particles as they scatter off the magnons (see Fig.~\ref{fig:spin-spin-drude}c). There is possibly a more direct signature of this crossover in {\it single-particle} spectral functions that can be measured, e.g. by tunneling experiments. In the Luttinger liquid setting, this allows the extraction of the charge Luttinger parameter, which is effectively doubled in the spin-incoherent regime relative  to its low-temperature, spin-coherent value. However these quantities are extremely difficult to compute via the TBA, as they involve form factors that do not admit the manifold simplifications of GHD. Furthermore, while natural in solid-state systems, they are less well-suited to the cold-atom setting. However, in optical lattice emulators of the Hubbard model,  quantum-gas microscopy techniques may allow the measurement of correlation functions and Drude weights.~\cite{PhysRevB.95.060406} Our work therefore leverages integrability to provide a complementary set of diagnostics for the crossover to those previously known. We expect that the basic structure is likely to survive, with minor modifications, in systems with weak  integrability breaking~\cite{friedman2019diffusive,2020arXiv200411030D} -- for instance, the $\delta$-function peak in the Drude response is broadened into a narrow Lorentzian with a decay time set by the scale of integrability breaking.	 Further investigations of the crossover regime in experimentally-relevant systems and observables seem warranted.

\subsection{Charge and energy transport}
	
	To complete the discussion, we now briefly summarize results for charge and energy transport away from half filling. Both can be understood analytically in most of the regimes identified above by using  appropriate expansions of the TBA and dressing equations; details are provided in Appendices~\ref{sec:app:TBAsol},~\ref{sec:app:largeM}, and~\ref{sec:app:largeU}, but we summarize the intuition behind the expansions for clarity. Formally, the TBA and dressing equations for the strings (i.e., the magnons, and the singlets) as presented them in Section~\ref{sec:TBA} are highly nonlocal in the species index, as they couple every species of string to every other species. This makes their solution computationally challenging even from a numerical perspective. However, they simplify in both the high-temperature and low-temperature regimes, as we now discuss. At high temperatures, long strings have appreciable filling, so that $n_{M|s}$ decays slowly for $M\to\infty$ (where $w$ or $s=uw$). In this limit, it is useful to recast the TBA and dressing equations into an alternative ``quasi-local'' form discussed in Appendix~\ref{sec:app:TBAsol}.  At low-temperatures, only short strings contribute and so it is safe to truncate the TBA equations even in their nonlocal form. In certain cases -- notably, in the spin-incoherent regime (iii) -- it is convenient to use a `hybrid' form of the TBA that invokes the nonlocal form for some species and the quasilocal form for others. (Heuristically, this can be understood by thinking of the magnons as being at high-temperature and tractable in the quasilocal form, and the singlets and $y$-particles being amenable to the low-temperature nonlocal description.)

	\begin{figure}
		\includegraphics[width=\linewidth]{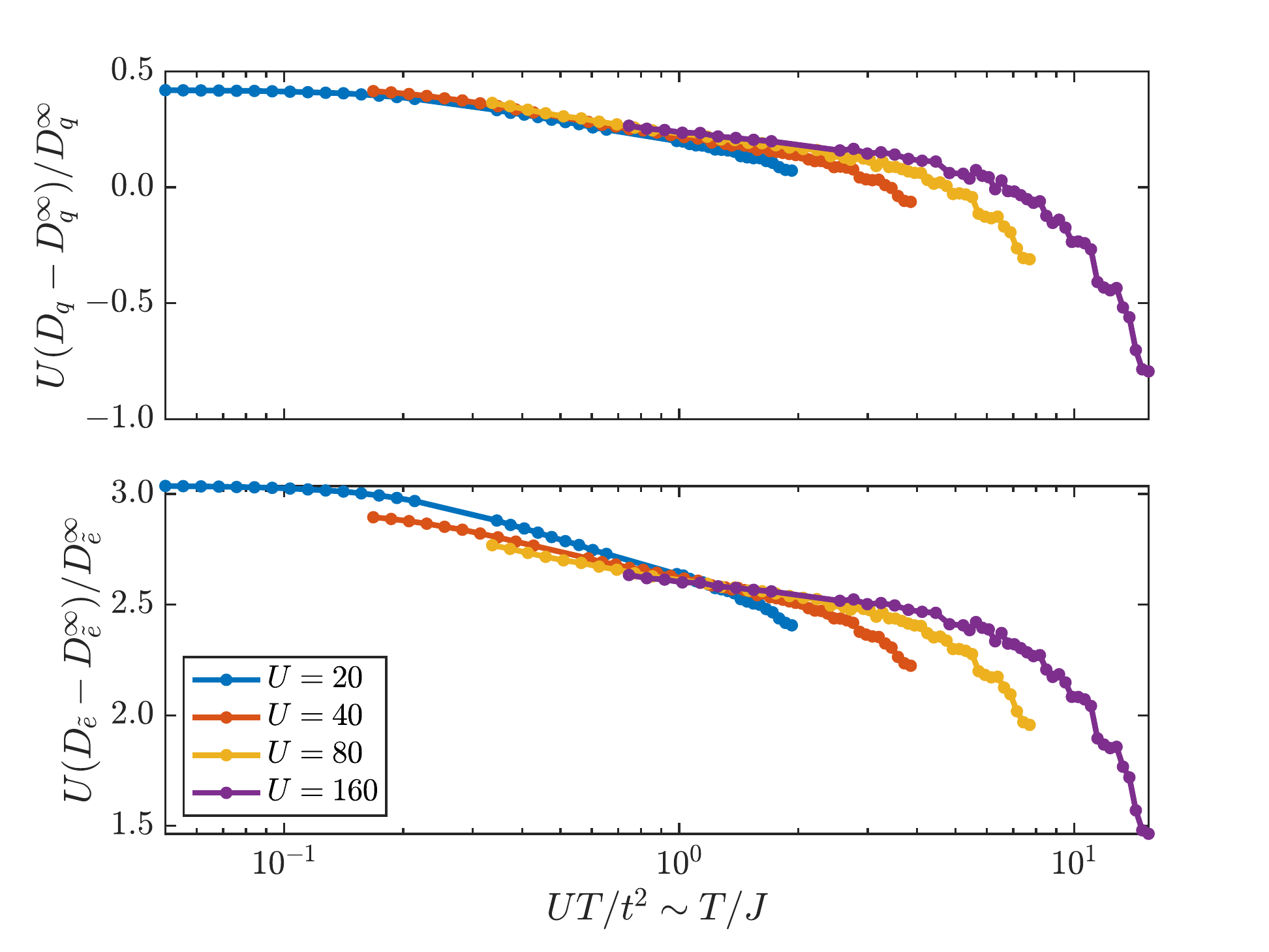}
		\caption{\label{fig:drude-subleading-crossover} The crossover between the spin-incoherent regime  (iii) and spin-coherent  regime (iv) is also visible in  subleading corrections to the charge and energy Dude weights. The plots show the magnitude of the relative correction to the Drude weight compared to the leading order expressions in $t/U$ ($D_q^\infty$ and $D_{\tilde{e}}^\infty$)  given by Eq.~\eqref{eq:large-U-Drude}. Apart from a tail at large $T$, which is due to the crossover to regime (ii), we see that the corrections indeed scale like $t/U$ and depend on the ratio $T/J$, signaling that their change is really a consequence of the (iii)-(iv) crossover. Numerical parameters: $h=0$ and $\langle n \rangle=0.3$.}
	\end{figure}

	In regime (i), we can can first perform a high-temperature expansion of the TBA equations and retain the first few terms to determine $n_a(u)$, and then expand in $t/U$ to solve for the density, the effective velocity and the dressed charges. Regimes (ii) and (iii) can be accessed instead by expanding directly in $t/U$. (Note however that since the $T\to\infty$ and $U\to\infty$ limits commute, regimes (i-iii) can be treated in a unified way.~\footnote{
		Indeed, in the TBA, the energy bandwidth of both $w$ and $uw$-strings is of order $J$, and the expansion is valid as long as $J/T\ll1$. The crossover between regimes (i) and (ii), is instead due to the chemical potential. Working at a fixed charge density and away from half filling requires $\mu=O(U)$, and therefore the crossover takes place when $\mu/T\sim 1$.  
	}
	The $t/U$ expansion breaks down in regime (iv), where we can, however, exploit the $T\to0$ limit in the presence of a finite magnetic field.
	
In all regimes, we find that the dominant contribution to the charge and energy Drude weights in the strong coupling limit is from the $y$-particles
	\begin{align}
		\label{eq:large-U-Drude}
		D_{\mathcal{O}} \simeq \frac{\beta}{2\pi} \sum_{a=y_\pm} \!\int\! du \, n_a(u) \bar{n}_a(u)  \left[\mathcal{O}_{a}^{\text{dr}}(u)\right]^2  \frac{\left[e'_a(u)\right]^2}{\left|k'_a(u)\right|}
	\end{align}
	where $\mathcal{O} =  q, \tilde{e}$ [Note that by the latter choice, we are focusing on the `reduced' energy Drude weight as discussed \revisionchange{at the end of Sec.~\ref{sec:GHD} and in Appendix~\ref{sec:app:TBAsol}}]. To obtain this strong-coupling expression for the Drude weight we used the fact that in the large-$U$ limit, and for $y$-particles, all  quantities apart from some dressed charges $\mathcal{O}_{a}^{\text{dr}}(u)$ are not dressed  to leading order in $t/U$, and applied the identity  $|k'_a(u)| =2 \pi \rho^t_a(u)$ {[See Appendix~\ref{sec:app:largeU}]}. Furthermore, it is understood that in \eqref{eq:large-U-Drude} the filling factor $n_\pm(u)$  is controlled by the bare energies.
	For energy transport $\tilde{e}_{\pm}^{\text{dr}}(u)$ is always dominated by its bare value $\tilde{e}_{\pm}(u)$ (Table~\ref{tab:TBA_spectrum}).
	To discuss the dressed electric charge, we need to distinguish regime (i), where
	\begin{equation}
		q^{\text{dr}}_{\pm} = \tanh\left(\beta\mu\right)<1
	\end{equation}
	with $\beta\mu$ implicitly determined by the filling, and regimes (ii), (iii), and (iv) where 
	\begin{equation}
		q^{\text{dr}}_{\pm}=1.
	\end{equation}

	Finally, the expression for the Drude weights can be further simplified in regimes (iii) and (iv), using the fact that $t\gg T $. In this situation, the Fermi factors $n_\pm(u)$ are step-like functions, jumping from $0$ to $1$ at two Fermi points $u_F$. Calling $v_F$ the bare (group) velocity at those points, we find
	\begin{equation}
	\label{eq:large-U-low-T-Drude}
		D_\mathcal{O} = \frac{v_F}{\pi} \mathcal{O}_F^2,
	\end{equation}
	where $\mathcal{O}_F$ is the operator evaluated at the Fermi points.
	
Expanding the TBA equations in regime (iv), we note that charge and energy transport do not change to leading order in $t/U$ during the the crossover from the spin-incoherent regime (iii) to the spin-coherent regime (iv). This was  postulated in the context of Luttinger liquid theory Ref.~\onlinecite{PhysRevB.72.125416} and was used to infer an effective theory of transport in the SILL. Using GHD, we have now verified that this statement is correct up to $t/U$ corrections (see Appendix~\ref{sec:app:largeU}).
Going beyond the leading terms, we also compute the exact charge and energy Drude weights by numerically solving the GHD equations; these are reported in Fig.~\ref{fig:drude-subleading-crossover}, which clearly shows that  these subleading corrections are sensitive to the crossover. An analytical estimate of the  corrections in regime (iii) can be found in Appendix~\ref{sec:app:largeU}.

	\section{\label{sec:halffilling}Transport at $h=0$ or $\mu=0$: KPZ universality and superdiffusion}
	
	We now turn to a generic feature of transport expected for all $t/U$, along special high-symmetry lines of the model. 
	As noted above, the Hubbard model hosts an $SU(2)_s$ symmetry whenever $h=0$ and an $SU(2)_\eta$ symmetry  when $\mu=0$. Along these high-symmetry lines, reasoning in analogy with the case of the isotropic Heisenberg (XXX) spin chain,~\cite{Ljubotina:2017aa, PhysRevLett.122.210602,PhysRevLett.121.230602,PhysRevLett.122.127202,PhysRevB.101.041411,2020arXiv200106432D,2020arXiv200313708D} we expect spin and/or charge transport respectively to be transported super-diffusively with length-time scaling governed by the Kardar-Parisi-Zhang  (KPZ) dynamical universality class~\cite{PhysRevLett.56.889}, meaning that
	\begin{align}
		\!\!\!\langle S^\mu(x,t) S^\mu(0,0) \rangle &= \frac{\chi_h}{[\lambda_{\text{KPZ}}^{(S)}t]^{2/3}} f_{\text{KPZ}}\left(\frac{x}{[\lambda_{\text{KPZ}}^{(S)}t]^{2/3}}\right), \label{eq:KPZ-scaling-S}\\
		\langle n(x,t) n(0,0) \rangle &=   \frac{\chi_\mu}{[\lambda_{\text{KPZ}}^{(\eta)}t]^{2/3}} f_{\text{KPZ}}\left(\frac{x}{[\lambda_{\text{KPZ}}^{(\eta)}t]^{2/3}}\right),\label{eq:KPZ-scaling-eta}
	\end{align}
	where $\chi_h$ and $\chi_\mu$ are respectively the spin and charge susceptibilities, $f_{\text{KPZ}}$ is a universal scaling function, and $\lambda_{\text{KPZ}}^{(S)}, \lambda_{\text{KPZ}}^{(\eta)}$ are characteristic energy scales for the KPZ dynamics.
	The possibility of superdiffusion in the Hubbard model was  first identified in Ref.~\onlinecite{PhysRevLett.121.230602}, that used bounding arguments to show that the diffusion constant diverged in the $h\to 0$ limit. However, a detailed analysis of  superdiffusive transport has not been previously attempted; also, the $SO(4)$ invariant point  $h=\mu=0$ has not been directly studied. Therefore, here we address these lacunae by providing arguments for KPZ scaling both along the high-symmetry lines and at the $SO(4)$ point, deploying both kinetic-theory approaches,~\cite{PhysRevLett.122.127202} and a classical analysis of soft gauge modes,~\cite{PhysRevB.101.041411,2020arXiv200106432D} before confirming our predictions using state-of-the-art numerical simulations using time-evolving Matrix-Product-Operators (MPOs).

	\subsection{Kinetic theory of superdiffusion\label{sec:kineticsuperdiff}}
	We begin our discussion of superdiffusion of charge and spin by incorporating diffusive corrections to the linearized GHD framework to demonstrate the divergence of the relevant diffusion constant, focusing for definiteness  on spin transport at $h=0$. 
	Following Refs.~\onlinecite{PhysRevLett.122.127202,2020arXiv200313708D} we estimate the effective spin diffusion constant at time $t$ as $\mathcal{D}_S(t) = \sum_{a} \mathcal{D}_a(t)$ where
	\begin{align} 
	\mathcal{D}_a(t) &=
	{\frac{ t}{4 \chi_h}}\int du \, \rho_a(u) \left[1-n_a(u)\right] \left[v^{\text{eff}}_a(u)\right]^2 \times\nonumber\\
	&~~~~~\sum_{j,k} \frac{1}{j!k!} \partial_{\tilde{\mu}}^j \partial_{\tilde{h}}^k \left[m^{\text{dr}}(\tilde \mu,\tilde h)\right]^2 \langle \tilde{\mu}^j \tilde{h}^k\rangle_t, \label{eq:diffusionkinetics}
	\end{align}
	where $\langle\cdot\rangle_t$ denotes the average up to time $t$ along the trajectory of the quasi-particle under consideration, and $\tilde{\mu}$ and $\tilde{h}$ are fluctuations in the effective chemical potential and effective magnetic field perceived by a propagating quasi-particle about their mean values  (respectively, $\mu/2$ and $0$). 	 [The expression~\eqref{eq:diffusionkinetics} for the diffusion constant can be obtained using the GHD by performing a gradient expansion, or by estimating the linear growth in time of the mean-square ``dipole moment''  $\langle (mx)^2\rangle$ of a spin excess initially localized at  the origin {(or equivalently, the spatial variance of the spin structure factor)}. The expression on the second line computes the additional dressed charge picked up by the quasiparticle as it propagates through a thermally fluctuating medium of other quasiparticles, order-by-order in fluctuations.]
	 
	 At long times, \eqref{eq:diffusionkinetics} only receives contributions from $(j,k)$ for which $\langle \tilde{\mu}^j \tilde{h}^k\rangle_t\propto1/t$, i.e. when $\langle \tilde{\mu}^j \tilde{h}^k\rangle_t$ is proportional to the inverse of the distance $l=|v^{\text{eff}}t|$ travelled by the particle.  Assuming Gaussian fluctuations of quasiparticle occupations (central limiting behavior) we find that
	\begin{align}
		\langle \tilde{h}^2 \rangle = \frac{1}{4\chi_h l}, ~~~~~\langle \tilde{\mu}^2 \rangle = \frac{1}{4\chi_\mu l},
	\end{align}
	where $\chi_h$ and $\chi_l$ are the magnetic and charge susceptibilities, with all higher moments scaling as higher inverse powers of $t$, and all cross terms vanishing due to the $S^z\to-S^z$ symmetry present for $h=0$.

	Hence, the diffusion constant is asymptotically given by the $t\to\infty$ limit of ~\eqref{eq:diffusionkinetics}, i.e. $\mathcal{D}_S = \sum_a \mathcal{D}^\infty_a$, with
	\begin{align}
	\mathcal{D}^\infty_a &\equiv\lim_{t\to \infty}\mathcal{D}_a(t) \nonumber\\&= \frac{1}{4 \chi_h} \int du\, \rho_a(u) \left[1-n_a(u)\right] \left|v^{\text{eff}}_a(u)\right| \nonumber\\
	&~~\left\{\frac{1}{4\chi_h} \partial_h^2 \left[m^{\text{dr}}_a(u)\right]^2 + \frac{1}{4\chi_\mu} \partial_\mu^2 \left[m^{\text{dr}}_a(u)\right]^2 \right\}.
	\end{align}
	For $h=0$, we see that the second term containing $\partial_\mu^2 \left[m^{\text{dr}}_a(u)\right]^2$ vanishes since $\left[m^{\text{dr}}_a(u)\right]^2$ is identically zero at $h=0$ due to the $SU(2)_s$ invariance, leaving us with the final result that
\begin{align}
	\!\!\!\mathcal{D}^\infty_a &= \int du \, \rho_a(u) \left[1-n_a(u)\right] \left|v^{\text{eff}}_a(u)\right| \frac{ \partial_h^2 \left[m^{\text{dr}}_a(u)\right]^2}{{16 \chi^2_h}}.
	\end{align}
	It remains to analyze this result for $h=0$.

	{We start by considering the case with $\mu<0$.
	As pointed out in Ref.~\onlinecite{PhysRevLett.121.230602}, the behavior at large $M$ can be understood from the asymptotic form~\cite{PhysRevB.65.165104}
 of $Y$ at large $M$	\begin{align}
		Y_{M|w}(u) &= \left[\frac{\sinh\left(f(u)+M\right)\beta h/2}{\sinh(\beta h/2)}\right]^2 -1,\\
		Y_{M|uw}(u) &= \left[\frac{\sinh\left(g(u)+M\right)\beta \mu}{\sinh(\beta \mu)}\right]^2 -1, \label{eq:YscalinglargeM}
 	\end{align}
 	for some ${O}(1)$ functions $f$ and $g$, which will generally depend on $\beta$, $h$ and $\mu$. Specifically, at $h=0$ and $\mu\neq0$ we  have for the magnons 
 	\begin{align}
 		Y_{M|w}(u) &\sim \left(f(u)+M\right)^2 -1,
 	\end{align}
 	while large singlet ($uw$) strings do not contribute to transport as their occupation is exponentially suppressed in $\beta\mu$.
 	Then, using the resulting that $m^{\text{dr}}_a=\partial_{\beta h} \log Y_a$, it follows that
 	\begin{equation}
 		m^{\text{dr}}_{M|w}(u) \sim \frac{1}{3} \left(f(u)+M\right)^2 \beta h.
 	\end{equation}
 	Combining this with the fact that \begin{equation}
	n_{M|w}\sim (f(u)+M)^{-2}, \label{eq:nMwscalinglargeM}
	\end{equation}
	 and that 
 	\begin{equation}
 		\int du\, \rho_{M|w}^t(u) \left|v^{\text{eff}}_{M|w}(u)\right|\sim \alpha/M^2, \label{eq:rhoMwscalinglargeM}
 	\end{equation}
 	for $h=0$ and large $M$ as in the XXX spin-chain (see Appendix~\ref{sec:app:largeM}), we have that $\mathcal{D}^\infty_{M|w}$ tends to a constant as $M\to \infty$. This produces a divergence in the spin diffusion constant. Since this mechanism is formally identically to the case of the XXX chain,  following the self-consistent argument in Ref.~\onlinecite{PhysRevLett.122.127202} we deduce that spin transport is superdiffusive with KPZ scaling exponent.
	
	We now focus on the case at $\mu=0$ (while keeping $h=0$ fixed), to examine if there is possible different structure to the divergence of $\mathcal{D}_S$ in this case. 	 	
 	When $\mu=0$, the magnons ($M|w$ strings) follow exactly the same scaling, and so would give rise to the same divergence in the diffusion constant; but now the singlets ($M|uw$ strings) are no longer exponentially suppressed and in principle could yield an additional  divergent contribution to $\mathcal{D}_S$. As it happens, however, using the fact (cf. ~\eqref{eq:YscalinglargeM}) that for large $M$
 	\begin{align}
 		Y_{M|uw}(u) &\sim \left(g(u)+M\right)^2 -1,
 	\end{align}
 	we find that $m^{\text{dr}}_{M|uw}\sim (\partial_{\beta h} g)/(g+M)$ with $\partial_{\beta h} g=0$ at $h=0$. We combine this with the analogs of Eqs.~\eqref{eq:nMwscalinglargeM} and ~\eqref{eq:rhoMwscalinglargeM} for the $uw$ strings (which have similar scaling at large $M$, see Appendix~\ref{sec:app:largeM}) to conclude that $\mathcal{D}_{M|uw}= {O}(M^{-6})$ and hence that $\sum_M {D}_{M|uw}$ converges. This strongly suggests that $z=3/2$ also for the $SO(4)$ case when $\mu=h=0$.  (Note that we obtain similar results by swapping the order of limits, suggesting that there is a well-defined $(\mu,h)\to(0,0)$ limit.) If we assert that  the scaling function is $f_{\text{KPZ}}$ also at $\mu =0$, it follows that $\lambda^{(S)}_{\text{KPZ}}(\mu=0)$ is smooth around $\mu=0$.
}

 	Finally, the above arguments about spin transport apply {\it mutatis mutandis}, for charge transport,  interchanging, e.g. the role of $h/2$ and $\mu$ and magnons and singlets ($w$- and $uw$-strings).
	
	\subsection{Soft gauge modes and KPZ universality from classical spin fluctuations\label{sec:softgauge}}
	While the kinetic approach  predicts KPZ-like exponents, it does not readily provide access to the KPZ scaling function.
	Therefore  we take the lead of recent work Ref.~\onlinecite{PhysRevB.101.041411} which proposed that super-diffusion emerges from the classical hydrodynamics of soft gauge modes. The Bethe ansatz  is built on a  choice of pseudovacuum, which in our case is the fermionic vacuum, and  quasi-particle excitations are characterized on top of this pseudovacuum. Evidently the fermionic vacuum preserves all microscopic symmetries. 	However, the choice of the pseudo-vacuum for the spin singlets necessarily breaks the $SU(2)_\eta$ symmetry, and similarly the choice of the  pseudo-vacuum for the magnons explicitly breaks  $SU(2)_s$ symmetry.~\cite{essler_2005}
	Soft gauge modes are dynamical space-dependent fluctuations of the pseudovacuum choice. In the XXX model, their classical dynamics -- governed by the Landau-Lifshitz equations --- was able to properly account for KPZ scaling. They have also been identified  identified as ``giant'' (large $M$) quasiparticles in recent work involving two of the present authors,~\cite{2020arXiv200313708D} providing a microscopic understanding of the emergence of superdiffusion via the GHD formalism; however at present we focus on the classical soft gauge approach.
	
	If either {$\mu$} or $h$ are non-zero, the discussion  proceeds identically as in Ref.~\onlinecite{PhysRevB.101.041411,2020arXiv200106432D}. We thus focus  on the case $h=0$ and $\mu=0$, where we have two soft gauge modes, associated to the breaking of $SU(2)_\eta$ and $SU(2)_s$ by our choice of pseudovacuum. We can parametrize this choice in terms of a pair of vector fields $\ve{\eta}(x,t)$ and $\ve{S}(x,t)$, that indicate respectively the expectation value of the operators $\ve{\eta}$ and $\ve{S}$ in the pseudo-vacuum.
	Working directly in the continuum limit, the dynamics of $\ve{S}(x,t)$ and $\ve{\eta}(x,t)$ will be described by a classical Hamiltonian $\mathcal{H}$, that produces Landau-Lifshitz dynamics, viz.
	\begin{align}
		\partial_t \ve{S}(x,t) &= \ve{S}\times \frac{\delta \mathcal{H}[\ve{S},\ve{\eta}]}{\delta\ve{S}(x)}\\
		\partial_t \ve{\eta}(x,t) &= \ve{\eta}\times \frac{\delta \mathcal{H}[\ve{S},\ve{\eta}]}{\delta\ve{\eta}(x)}.
	\end{align}
	On symmetry grounds, we consider the most general $\mathcal{H}$ that is invariant under all the relevant symmetries, and particularly under the transformations generated by the  $su(2)_s\oplus su(2)_\eta$ algebra. This is the symmetry class of two independent spin chains, invariant under independent rotations of the spin in each chain. This means that the only terms that can appear in the Hamiltonian are rotational scalar intra-chain couplings, and inter-chain scalar-scalar couplings.  
	
	Focusing first on the intra-chain coupling, the most relevant terms (i.e. those with the lowest number of derivatives) in the equations of motion are then given by
	\begin{align}
		\partial_t \ve{S}(x,t) &= J_S \ve{S}\times \partial_x^2 \ve{S}\\
		\partial_t \ve{\eta}(x,t) &= J_\eta \ve{\eta}\times \partial_x^2 \ve{\eta}.
	\end{align}
	These equation becomes particularly simple once expressed using Frenet-Serret variables~\cite{LAKSHMANAN1976577}
	\begin{align}
		\kappa_S &= \sqrt{\left(\partial_x \ve{S}\right)^2}\\
		\tau_S & = \frac{1}{\kappa_S^2} \ve{S}\cdot\left(\partial_x \ve{S} \times \partial_x^2 \ve{S}\right)		
	\end{align}
	and similarly for $\ve{\eta}$. In term of these, we have
	\begin{align}
		\partial_t \kappa_a^2 &= -J_a \partial_x\left(\kappa_a^2\tau_a\right)\\
		\partial_t \tau_a &= -J_a \partial_x \left(\tau_a^2 - \kappa_a^2/2 - \partial_x^2(\kappa_a)/\kappa_a\right)
	\end{align}
	with $a=\eta,~S$. Upon coarse-graining over a length scale $l$, $\kappa_a^2\sim1/l^2$ 
	and  is  transported ballistically; the latter follows from the fact that $\kappa_a^2$ is proportional to the energy density and is hence ballistic because of integrability. Meanwhile, as we argue self-consistently below $\tau$ will be transported super-diffusively. Therefore, the two equations effectively dynamically decouple.~\cite{PhysRevB.101.041411,2020arXiv200106432D}	
	Focusing on the second equation, we insert {a phenomenological} diffusion coefficients $D_a$ and white noise terms $\xi_a$, and thereby obtain a pair of uncoupled noisy Burger equations for the $\tau_a$,
	\begin{align}
		\partial_t \tau_a = J_a \partial_x \left(-\tau_a^2 +D_a \partial_x \tau_a + \xi_a\right).
	\end{align}
	The solutions of these independent equations each obey KPZ scaling. From the perspective of the Burgers equations, the only relevant terms we can write that couple the two equations are of the form $\partial_x(\tau_S\tau_\eta)$. Although this term could under special cases produce different scaling exponents (see e.g. Ref.~\onlinecite{PhysRevLett.69.929}) and could more generally produce a renormalization of the KPZ scaling function,~\cite{Spohn2014} it is not  {\it a priori} obvious if such a coupling can arise under the restriction of the $SU(2)_\eta\times SU(2)_s$ symmetry and from local lattice Hamiltonian dynamics. Indeed,  we will argue below that regular scalar-scalar couplings cannot give rise to a term of this form. This then leads us to conclude that the scaling of spin-spin and charge-charge correlators is strictly KPZ also at the $SO(4)$ point (though there may be significant finite-size effects relative to the single-KPZ case since there are additional irrelevant `interchain' couplings that must flow to zero before the two Burgers equations decouple).
	
	The terms in the continuum that admit an obvious regularization on the lattice are polynomials of $\ve{S}$, $\ve{\eta}$, $\partial_x^n\ve{S}$ and $\partial_x^n\ve{\eta}$. Rotational scalars can then be constructed either by taking the scalar product of two derivatives of the vector field (e.g. $\partial_x^n\ve{S}\cdot \partial_x^m\ve{S}$), or as triple products (e.g. $\partial_x^n\ve{S}\cdot \left(\partial_x^m\ve{S}\times\partial_x^l\ve{S}\right)$).
	
	In order to show that such terms cannot produce a $\tau_S\tau_\eta$ coupling, we recapitulate the Frenet-Serret formalism. Focusing for specificity on the spin dynamics, the key idea is fix a space and time dependent frame (Frenet-Serret frame) characterized by the $3$ unit vectors $\boldsymbol{e}_{S,1}= \ve{S}$, $\boldsymbol{e}_{S,2}= (\partial_x\ve{S})/\kappa_S$ and $\boldsymbol{e}_{S,3}$. Since the  frame is space-dependent, its spatial variation can be described via the pseudo-vector $\ve{\Omega}_{S}(x,t)$, i.e. $\partial_x \ve{e}_{S,j}=\ve{\Omega}_{S}\times \ve{e}_{S,j}$. Similarly the time-variation of the frame can be described in terms of its angular velocity $\ve{\omega}_{S}(x,t)$. From these two pseudo-vectors we can describe any derivative of a vector  $\ve{v}$ as
	\begin{align}
		\partial_x \ve{v} &= \partial^{(\text{FS})}_x \ve{v} + \ve{\Omega}_{S} \times \ve{v}\\
		\partial_t \ve{v} &= \partial^{(\text{FS})}_t \ve{v} + \ve{\omega}_{S} \times \ve{v},
	\end{align}
	with $\partial^{(\text{FS})}$ denoting the partial derivative in the Frenet-Serret frame.
	To determine the dynamics of $\tau_S$, we exploit that~\cite{LAKSHMANAN1976577} in the Frenet-Serret frame $\ve{\Omega}_{S}=(\tau_S,0,\kappa_S)$ and that~\cite{LAKSHMANAN1976577,2020arXiv200106432D} $\partial_x \tau_S = \partial\omega_{S,1}-\kappa_S \omega_{S,2}$. Finally $\ve{\omega}_S$ is determined by the classical Hamiltonian dynamics of the system as
	\begin{align}
		\omega_{S,i} &=  -\left(\frac{\delta \mathcal{H}[\ve{S},\ve{\eta}]}{\delta\ve{S}(x)}\right)_i\quad&i=2,3\\
		\omega_{S,1} &= \frac{\partial_x^{(\text{FS})}\omega_{S,2}+\tau_S\omega_{S,3}}{\kappa_S}.
	\end{align}
	Therefore, even if the Hamiltonian $\mathcal{H}$ includes terms of the form $F\left(\ve{S},\partial_x\ve{S},\cdots\right)G\left(\ve{\eta},\partial_x\ve{\eta},\cdots\right)$, the final equation of motion for spin torsion will be of the form
	\begin{align}
		\partial_t \tau_S &= \partial_x \left(J_1[\kappa_S,\tau_S]G\left(\ve{\eta},\partial_x\ve{\eta},\cdots\right)\right) \\&~~+ \partial_x \left(J_2[\kappa_S,\tau_S]G\left(\ve{\eta},\partial_x\ve{\eta},\cdots\right)\right)\nonumber
	\end{align}
	for some functionals $J_1$ and $J_2$ which can be computed from $F$. Crucially, the function $G$ is left unaltered in computing the equation of motion for $\tau_S$. To obtain $\partial_t \tau_S=\cdots+ \partial_x(\tau_S\tau_\eta)$ then we would need $G=\tau_\eta$, which is not possible to achieve using only polynomials of the derivatives of $\ve{\eta}$, but would require a continuum Hamiltonian which would not admit a trivial  lattice regularization.
	
Thus, we conclude that no lattice-regularizable classical Hamiltonian can produce a coupling between the Burgers equations for $\tau_S$ and $\tau_\eta$ that is relevant under KPZ scaling. As a consequence, the scaling of spin-spin and charge-charge correlators is of the ``$(\text{KPZ})^2$'' form given in Eqs.~\eqref{eq:KPZ-scaling-S} and \eqref{eq:KPZ-scaling-eta}. 	

\subsection{Numerical Simulations}
	We confirm the double-KPZ scaling scenario presented above by means of state-of-the-art time-evolving block-decimation (TEBD) numerical simulations using the matrix product operator (MPO) formalism. We focus on spin dynamics at $h=\mu=0$ and $T=\infty$, and compute the dynamic correlator $\langle S^z(x\revisionchange{_j}, t)S^z(0, 0)\rangle$, where $j$ indexes  the sites. To do this, we represent $S^z(0, 0)$ as a bond-dimension-$1$ MPO, which we subsequently evolve in the Heisenberg picture using TEBD techniques. All evolutions are done with a fourth-order Trotter step of size $\delta t = 0.2$. Truncations are done initially with a fixed discarded weight $\varepsilon = 10^{-8}$ and a growing bond dimension --- however, once the bond dimension surpasses a threshold, subsequent truncations keep at most  $\chi_{\text{max}}$ states, with $\chi_{\text{max}}$ ranging from $256$ to $1024$. Our conclusions are quantitatively consistent across bond dimensions.

	\begin{figure}[t!]
		\includegraphics[width=0.95\linewidth]{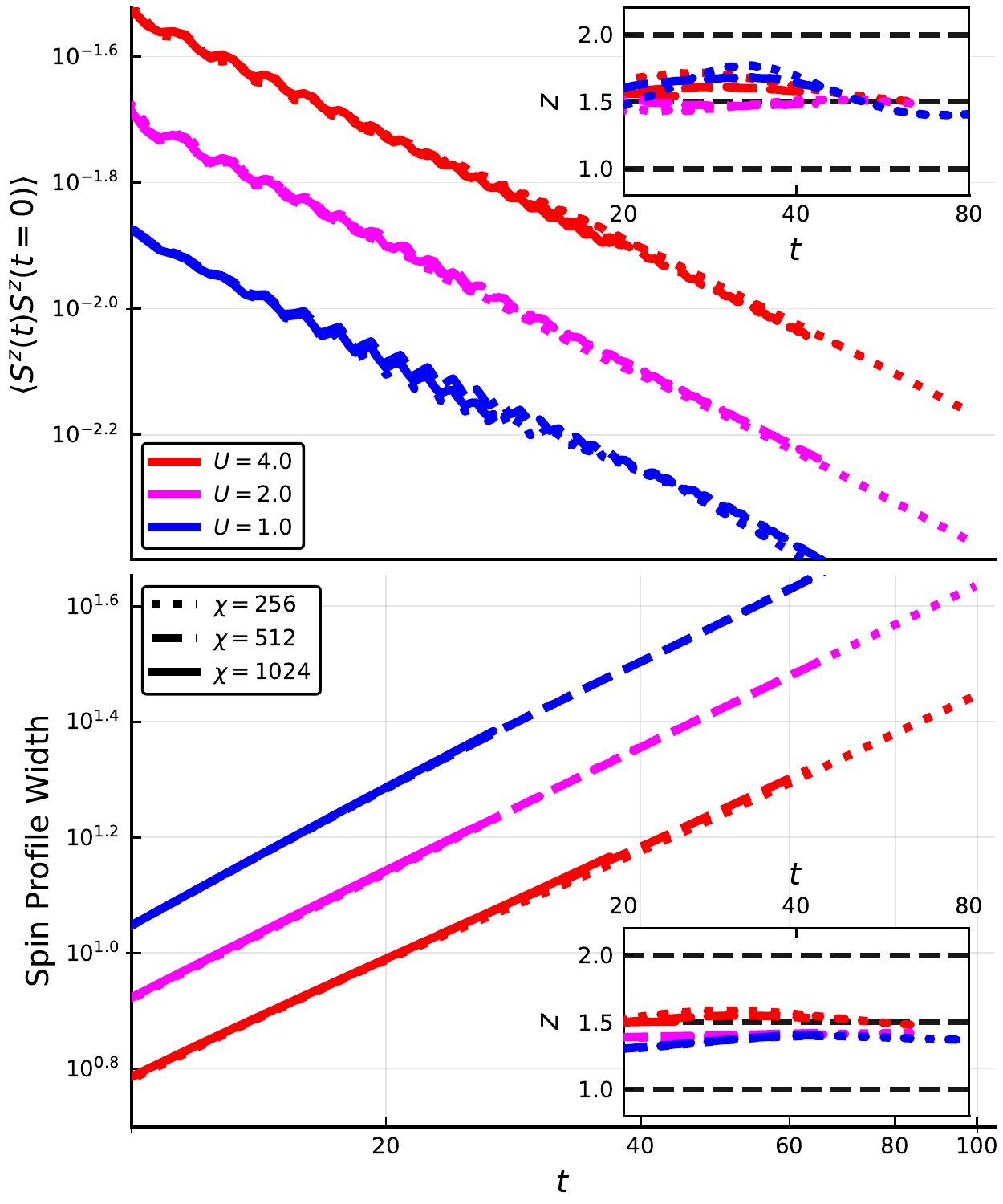}
		\caption{\label{fig:TEBD-Sz-combined} TEBD data for the Hubbard model at $h=\mu=0$ and $U/t=1, 2, 4$ and for various maximum-bond-dimensions $\chi_{\text{max}} = 256, 512, 1024$. From the return probability (top), and from the profile width (bottom), we have fit the dynamical exponent $z$ (insets) as described in the main text.}
	\end{figure}

	Once we have $S^z(0, t)$, we exploit translational invariance to access $S^z(x\revisionchange{_j}, t)$, at which point the correlator can be easily computed.
	To reduce the error coming from the trotterization and the SVD truncation, we exploit the sum rule
	\begin{equation}
		\sum_j \langle S^z(x\revisionchange{_j}, t)S^z(0, 0)\rangle = \chi_h,
	\end{equation}
	to correctly normalize $\langle S^z(x\revisionchange{_j}, t)S^z(0, 0)\rangle$ at each time $t$.
	
	We consider two different methods to extract the dynamical exponent $z$ and $\lambda_{\text{KPZ}}^{(S)}$. First, we analyze the return probability $\langle S^z(0, t)S^z(0, 0)\rangle$ as a function of time; second, we analyse the growth of the profile width
	\begin{equation}
		\sqrt{\sum_j x_j^2\langle S^z(x_j, t)S^z(0, 0)\rangle}.
	\end{equation}
	with $j$ indexing the sites and $x_j$ measured in units of lattice spacings.
	At each time $t$, we take a window of size $\Delta \log t = 1$ centered around $t$ and fit the outcome of the TEBD simulations within that time window to the predicted KPZ scaling form. The results we obtain are presented in Fig.~\ref{fig:TEBD-Sz-combined} for $U/t=1, 2$, and $4$. Our analysis suggests that $z$ and $\lambda_{\text{KPZ}}^{(S)}$ converge more quickly in time when the return probability is analysed. Both approaches, however, seem to be compatible with $z=3/2$ at sufficiently late times.

    Note that, especially at $U/t=4$, the fit for $z$ seems to converge rather quickly, suggesting that the exponent $z$ could be accessed in the timescale of a typical cold-atom experiment. To further highlight the accessibility to quantum microscope experiments, we used TEBD to analyze finite size effects, finite temperature effects, and the effects of small $SU(2)_s$ symmetry breaking field $h$ and $SU(2)_{\eta}$ symmetry breaking chemical potential $\mu$ around the $U/t=4$ point.
    
For short chains, the clearest signature of $z=3/2$ scaling is in the autocorrelator for a spin near the middle of the chain, which shows $t^{-2/3}$ scaling before eventually saturating to $0.5/L$ in a time that scales like $L^{3/2}$ (Fig.~\ref{fig:TEBD-Sz-L}).
    \begin{figure}[t!]
        \includegraphics[width=0.95\linewidth]{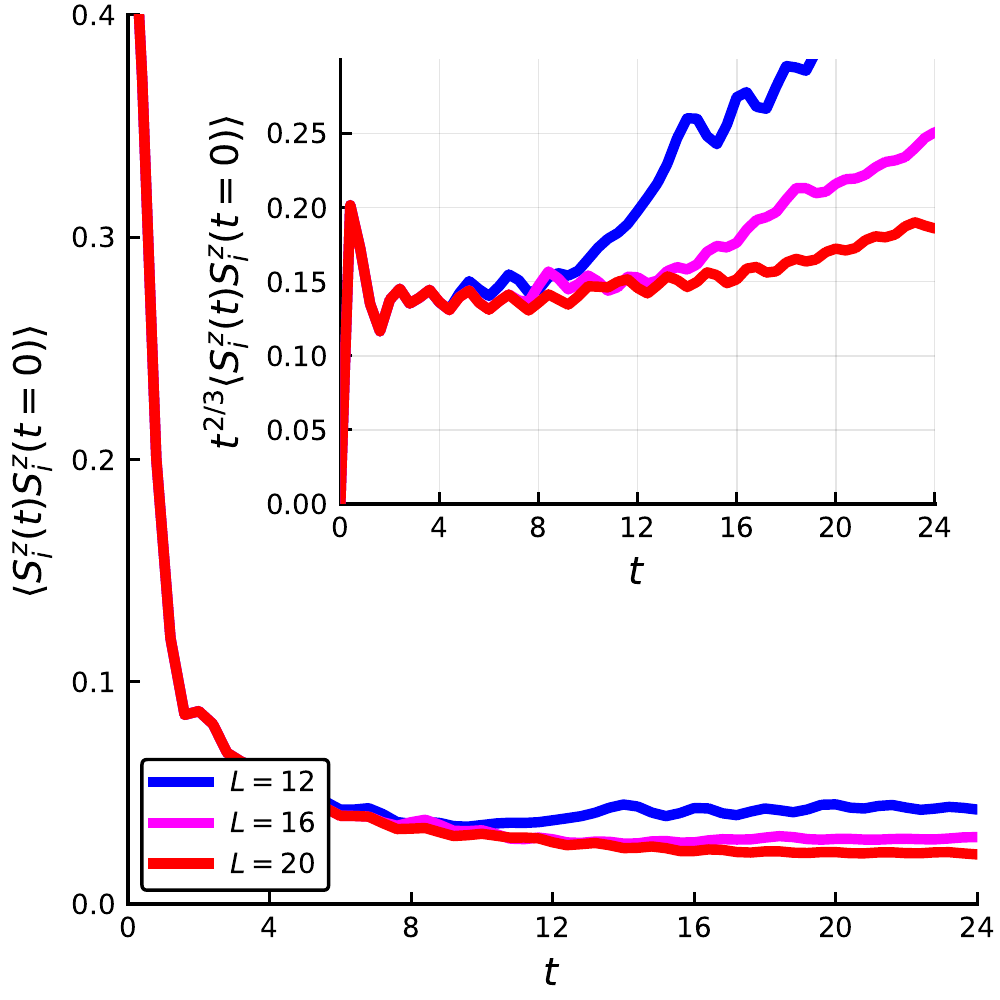}
        \caption{\label{fig:TEBD-Sz-L} Signatures of superdiffusion  in small Hubbard chains directly accessible in current quantum microscope experiments. TEBD data for the Hubbard model at $U/t=4$ for finite chains of size $L=12, 16, 20$. The return probability saturates due to the finite size effects, cutting off the $t^{-2/3}$ scaling.}
    \end{figure}
    Upon decreasing the temperature from $T=\infty$ to $T=2t$, the $z=3/2$ scaling remains (Fig.~\ref{fig:TEBD-Sz-T}).
    \begin{figure}[t!]
        \includegraphics[width=0.95\linewidth]{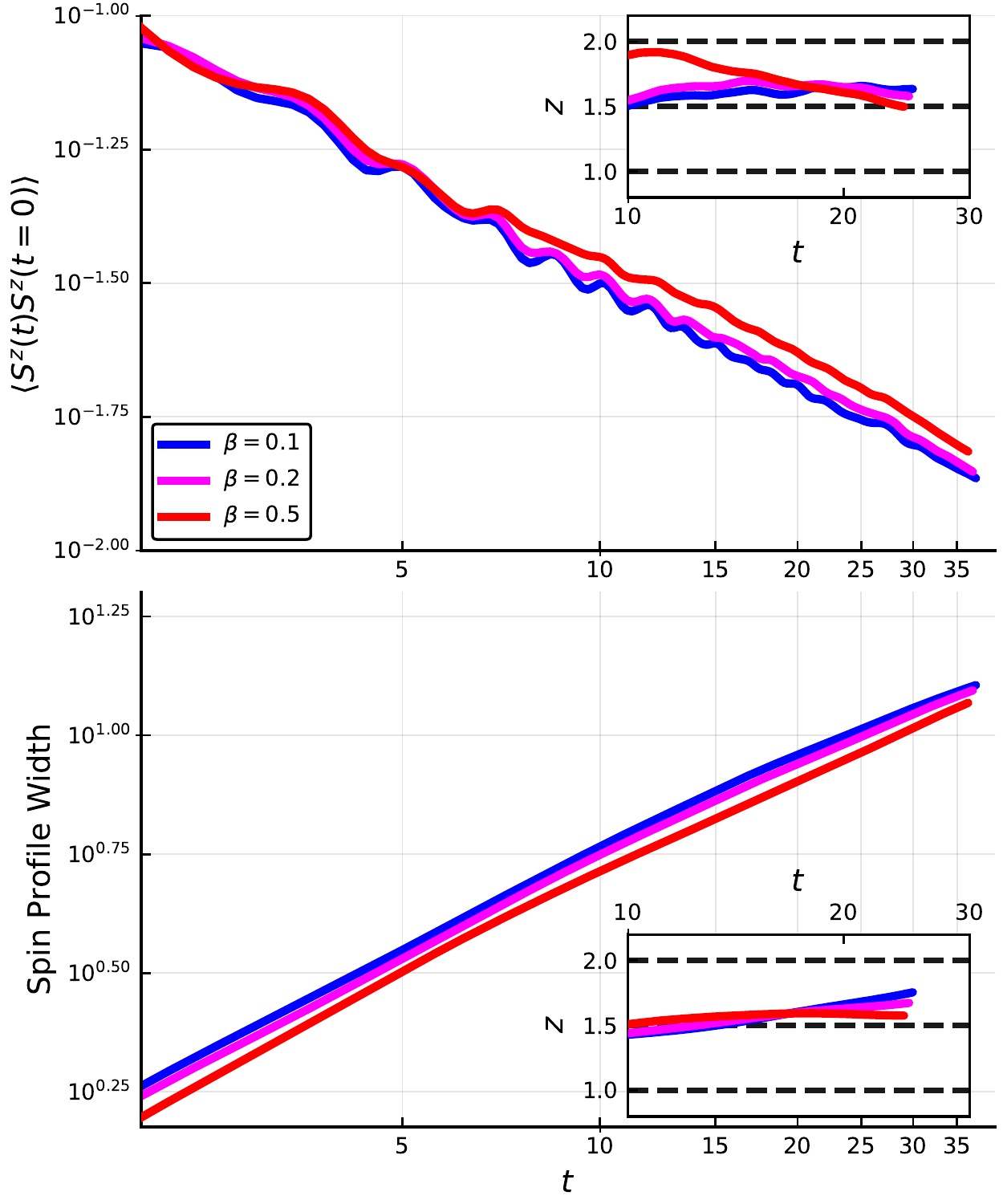}
        \caption{\label{fig:TEBD-Sz-T} Superdiffusion at finite temperature. TEBD data for the Hubbard model at $U/t=4$ for $\beta t = 0.1, 0.2, 0.5$.}
    \end{figure}

    Moving away from the $SU(2)_s$ symmetric line by adding a small field $h$, the spin structure factor $\langle S^z(x\revisionchange{_j}, t)S^z(0, 0)\rangle_c$ immediately develops ballistically moving peaks corresponding to light magnons (Fig.~\ref{fig:TEBD-Sz-h}), which coexist with a superdiffusive peak at $x=0$, which we expect~\cite{Gopalakrishnan16250} will cross over to a ballistic scaling at times $\gg h^{-3}$.
    \revisionchange{
    	Indeed, as $h\to 0$ the spin Drude weight $D_m$ scales as as $h^2 |\log h|$, since the contributions of $M|w$ strings to $D_m$ scale like $h^2/M$ up to $M\sim 1/h$, beyond which the $n_{M|w}$ is exponentially suppressed (see  Ref.~\onlinecite{PhysRevLett.121.230602} for a more detailed discussion). Thus, in the ballistic regime, the spatial variance of the spin profile is proportional to $D t^2 \sim h^2 |\log h| t^2$ at long times. At short times, instead, $h$ is effectively $0$ and the variance of the spatial profile is given by the KPZ scaling form, i.e. it is proportional to $t^{4/3}$. Therefore, the crossover from anomalous diffusion to ballistic transport takes place when these two lengthscales becomes comparable, i.e. on a timescale $t^*\sim h^{-3}$ (up to logarithmic corrections).
	}
	
    \begin{figure}[t!]
        \includegraphics[width=0.95\linewidth]{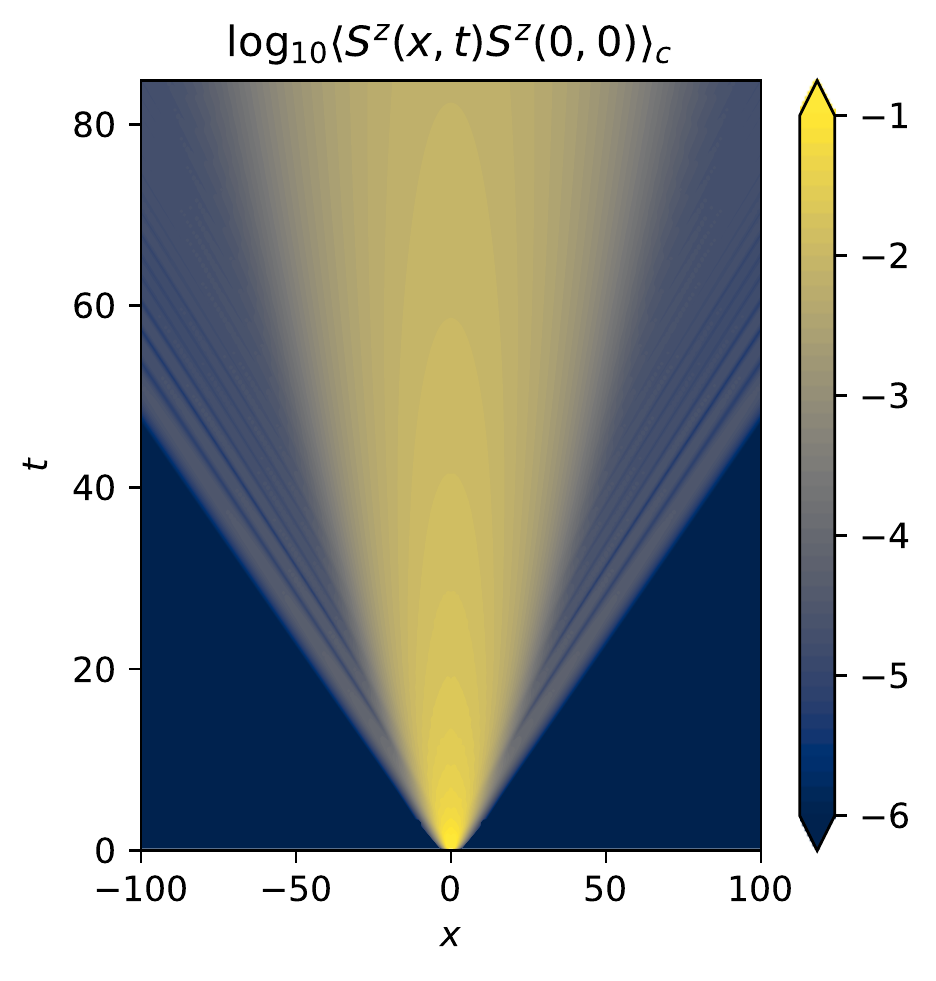}
        \caption{\label{fig:TEBD-Sz-h} TEBD data for the Hubbard model at $h=0.1, \mu=0$ and $U/t=4$. Ballistically moving peaks coexist with a superdiffusive central peak that lasts for a time scale $\sim h^{-3}$.}
    \end{figure}
    
    	\begin{figure*}[t!]
		\includegraphics[width=\linewidth]{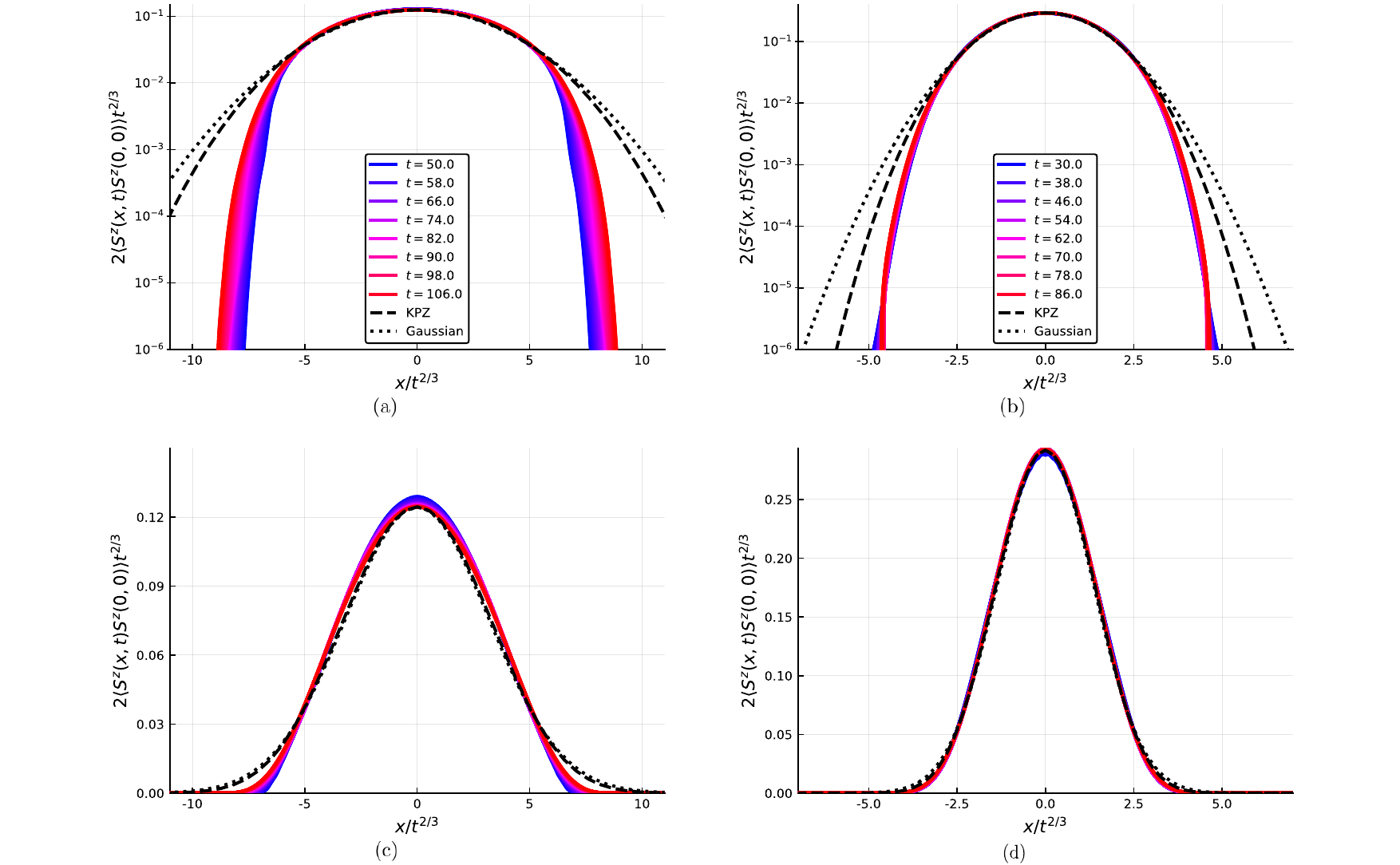}
		\caption{\label{fig:profile-collapse} (a) Collapsed profile of $\langle S^z(x\revisionchange{_j}, t)S^z(0, 0)\rangle$ for $U/t=1$ assuming $z=3/2$. We can observe that the tails of the correlators did not converge yet up to the maximum time we were able to reach in our simulations. (b) Same as panel (a), but for $U/t=4$. The drift of the profile with time is less pronounced than in (a), making it harder to understand if the profile is converged in $t$. If convergence has been reached, this would indicate that the profile are described by $f_{\text{KPZ}}$, in contrast with our previous analysis. Panel (c) and (d) report the same data of (a) and (b) respectively, but in a linear scale.}
	\end{figure*}
    
    {In the parameter regime of these numerics, adding a small chemical potential $\mu$ also has no effect on the $z=3/2$ scaling of $\langle S^z(x\revisionchange{_j}, t)S^z(0, 0)\rangle_c$}; this is consistent with our analysis in Sec.~\ref{sec:kineticsuperdiff} above, since as we have discussed $\mu$ breaks the $SU(2)_\eta$ symmetry but preserves $SU(2)_s$.
    Thus we expect that the $z=3/2$ scaling should be accessible to currently available experimental platforms (see e.g. Ref.~\onlinecite{Vijayan186}), with the biggest limitation being imposed by the finite length of the chains.

{Finally, we analyze the full profile of $\langle S^z(x\revisionchange{_j}, t)S^z(0, 0)\rangle$ at different times (Fig.~\ref{fig:profile-collapse}), assuming $z=3/2$, with the goal of determining if the scaling function is of the KPZ form $f_{\text{KPZ}}$, as indicated by our soft gauge mode-treatment. 
	At the latest times for which our TEBD truncation errors are controlled, the profiles are not converged in the tails; these tails seem to fall off faster than $f_{\mathrm{KPZ}}$. Therefore we cannot definitively conclude that the scaling function is of KPZ form. Note that coupled noisy Burgers equations can give rise to non-KPZ scaling functions consistent with the exponent $z = 3/2$.\cite{PhysRevLett.69.929} Our numerical results do not rule out this possibility. We emphasize, however, that we see strong similarities between our numerical data for the scaling function in the Hubbard model at $U/t=4$ and our data for the Heisenberg spin chain, shown in Fig.~\ref{fig:profile-collapse-heisenberg}, computed using the same TEBD approach. Whether convergence to the KPZ scaling function appears on larger time scales is an interesting question for future work.}

\begin{figure}[t!]
\includegraphics[width=0.95\linewidth]{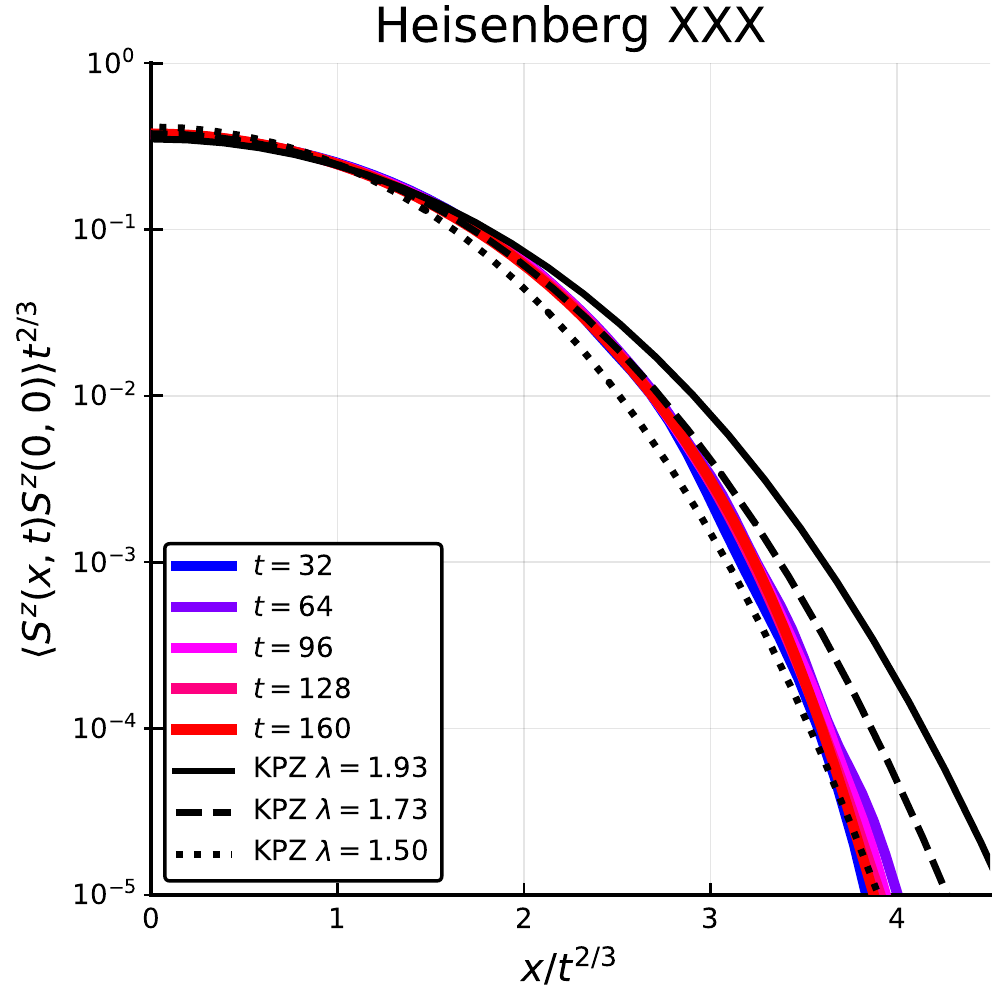}
\caption{\label{fig:profile-collapse-heisenberg} Collapsed profile of $\langle S^z(x\revisionchange{_j}, t)S^z(0, 0)\rangle$ for the Heisenberg XXX chain assuming $z=3/2$, computed using TEBD with bond dimension $\chi=512$. Significant deviations in the tails from $f_{\text{KPZ}}$ persist to accessible timescales.}
\end{figure}

	\subsection{Energy transport at the $SO(4)$ point} 
		Finally, for completeness we briefly discuss the ballistic transport of energy at half filling and in zero field. This was previously studied in Refs.~\onlinecite{PhysRevLett.117.116401,Karrasch2017,PhysRevB.96.081118}, where  the energy Drude weight for the Hubbard model at half-filling and zero magnetization were computed, uncovering a rich structure. We once again focus on the strong coupling limit, where the energy Drude weight dependence on temperature is the richest, and discuss transport in its various regimes. While the expression~\eqref{eq:large-U-Drude} still gives the leading contribution to the energy Drude weight in regime (i) [via ~\eqref{eq:large-U-Drude}], in regimes (ii)-(iv) $\bar{n}_\pm(u)$ is exponentially suppressed in $U/T$ for half filling and hence the $y$-particles do not contribute significantly to the energy Drude response. Similarly, the total density of singlets $\rho^t_{M|uw}$ at half filling is suppressed by the same factor of $U/T$. Therefore, the dominant contribution to the thermal (energy) Drude weight comes from magnons  ($w$-strings) in regime (ii), (iii) and (iv). Accordingly, we have (by similar manipulations as those that led to~\eqref{eq:large-U-Drude}) that the energy Drude weight in regimes (ii), (iii), and (iv) is given by
\begin{align}
		\label{eq:energyDrudehalf}
		D_e \simeq \frac{\beta}{2\pi} \!\sum_{a=M|w} \!\int \!\!du \, n_{a}(u) \bar{n}_{a}(u) \left[e_{a}^{\text{dr}}(u)\right]^2  \frac{\left[\left(e_{a}'\right)^{\text{dr}}(u)\right]^2}{\left|\left(k'_{a}\right)^{\text{dr}}(u)\right|}.
	\end{align}

	In regimes (ii) and (iii), the energy Drude weight can be explicitly computed as (see Appendix~\ref{sec:app:largeU}), leading to
	\begin{align}
	\label{eq:energyDrudehalf-explicit}
		D_e \simeq  20.05~\frac{\beta t^{9}}{U^4}
	\end{align}
	which monotonically increases as $T$ decreases towards regime (iv).
	{Finally, deep in regime (iv) as $T\to0$, the energy Drude weight is dominated by a Fermi point of the $1|w$-strings (``elementary'' magnons) at infinite rapidity . As shown in Appendix~\ref{sec:app:largeU}, this contribution is
	\begin{equation}
		D_e \simeq \frac{\pi^2 t^3 T^2}{3U},
	\end{equation}
	i.e.  $D_e$ monotonically decreases with the temperature.
	}
	
	Combining the above results, we can qualitatively understand the behavior of $D_e$ presented in Refs.~\onlinecite{Karrasch2017,PhysRevB.96.081118}. $D_e$ starts of order $\beta t^5$ at large $T$ in regime (i). Decreasing $T$, when $ U/T\lesssim 1$ $D_e$  rapidly deceases to become suppressed as $\beta t^9/U^4$ in regimes (ii-iii). As $T$ decreases again, approaching regime (iv), $D_e$ slowly increases monotonically. As we know that $D_e$ is decreasing towards $0$ in regime (iv), it must attain a maximum at the (iii)-(iv) crossover. Away from half-filling, we linked similar features in spin transport at this scale  with the crossover between spin-coherent and spin-incoherent behavior.  In  the half-filled case, this can be identified as a `Hubbard to Heisenberg crossover' linked to the freeze-out of charge fluctuations [see Refs.~\onlinecite{PhysRevLett.117.116401,Karrasch2017,PhysRevB.96.081118} for a discussion].

	\section{\label{sec:conclusions}Concluding Remarks}

In this work we revisited transport in the paradigmatic Hubbard model, in one dimension, in light of recent developments in understanding transport in integrable systems using generalized hydrodynamics. The GHD framework allowed us to capture the crossover between the low temperature Luttinger liquid, the intermediate-temperature spin-incoherent Luttinger liquid, and the high temperature regime. Away from half filling (zero magnetic field), charge (spin) transport is primarily ballistic. We explored the crossovers between various ballistic regimes, focusing on the Drude weight and the dynamic structure factor as diagnostics. The sharpest crossovers away from the $SO(4)$ point are in spin transport, as one might expect. In all regimes, when $\mu\neq0$, spin transport is dominated by the fast-moving $y$-particles, whose dressed magnetization is sensitive to the crossover. The long time spin structure factor $\mathcal{S}_m(x,t)$ reflects the spin-incoherent to spin-coherent crossover more directly. In the former case, $\mathcal{S}_m(x,t)$ displays a hierarchy of peaks of increasing height as $x/t$ decreases. Instead, in the latter,  $\mathcal{S}_m(x,t)$ is dominated by $1|w$-magnons. Less pronounced signatures are seen in energy and charge transport away from half filling.
Finally, we turned to the case of half filling and/or zero magnetic field: in this limit, ballistic transport vanishes, and instead one has superdiffusive charge and spin transport, which we argued belongs to the KPZ universality class. We presented extensive numerical evidence for $z=3/2$ dynamical scaling.

We close with two remarks. First, the most striking qualitative phenomena we have found (such as charge and spin superdiffusion) are observable in quite small systems of $L \alt 22$ at times $t \alt 20$. These can easily be realized in experiments using quantum gas microscopes, as well as other existing or near-term experiments. The GHD framework---built on physically reasonable but unrigorous assumptions---makes exact, zero-parameter predictions for such experiments, which it is important to test. Second, one can regard the quasiparticle picture presented here as the starting point for a broader analysis of high-temperature transport in the Hubbard model, perturbed slightly away from integrability. Integrability-breaking creates decay channels for all the quasiparticle types we have considered; however, owing to their very different kinematics, we expect a family of well-separated quasiparticle lifetimes, and consequently a sequence of dynamical crossovers that persist in the nonintegrable limit as well. 
	
\begin{acknowledgments}
{We thank J. De Nardis, F. Essler, E. Ilievski, M. Zwierlein, and especially S. Roy for useful discussions}. S.G., R.V. and B.W. also thank J. De Nardis and E. Ilievski for collaboration on related works. We thank V. Bulchandani for useful comments on the manuscript. 
S.G. acknowledges support from NSF Grant No. DMR-1653271 and PSC-CUNY Grant No. 63632-00 51. S.A.P. acknowledges support from EPSRC Grant EP/S020527/1. R.V.  acknowledges support from the US Department of Energy, Office of Science, Basic Energy Sciences, under Early Career Award No. DE-SC0019168,  and the Alfred P. Sloan Foundation through a Sloan Research Fellowship. 

\end{acknowledgments}

	\begin{appendix}
	\section{\label{sec:app:TBA}Review of Thermodynamic Bethe Ansatz for the Hubbard Model}
	
	Here we give a brief summary of the salient features  of the TBA for the one-dimensional Hubbard model, providing more details of the results quoted in the main text. Much of this material may be found in classic monographs [Refs.~\onlinecite{takahashi_1999,essler_2005}];   to our knowledge its first application in conjunction with the GHD formalism is in Ref.~\onlinecite{PhysRevB.96.081118}.

	\begin{table*}[]
		\caption{\label{tab:TBA_spectrum}TBA Spectrum for the Hubbard model. As is customary, we employ units in which the hopping strength, $t=1$. Here, $\tilde{e}_a(u)$ is the dressed energy without the contribution from the chemical potential and magnetization terms in the Hamiltonian (i.e. for $\tilde{H} = \hat{T} + \hat{V}$). The full dressed energy for $H$ is ${e}_a(u) = \tilde{e}_a(u) - \mu q_a - h m_a $.} 
		\begin{tabular}{|c|c|c|c|c|c|c|}
			\hline
			\textbf{Species $\boldsymbol{a}$}       & \textbf{$\boldsymbol{u_a}$ domain}      & {$\boldsymbol{\sigma_a}$}  & $\boldsymbol{q_a}$  & $\boldsymbol{m_a}$           & $\boldsymbol{k'(u)}$                       & $\boldsymbol{\tilde{e}_a(u)}$      \\ \hline
			$y$ ($\pm$-branch)   & $[-1,1]$             & $\mp1$    & $1$  & $\frac{1}{2}$            & $\mp \frac{1}{\sqrt{1-u^2}}$      & $\pm 2\sqrt{1-u^2}-\frac{U}{2}$   \\ \hline 		$M|uw$ &
			$\mathbb{R}$ &
			$-1$ &  $2M$  & $0$ &
			$-\frac{1}{\sqrt{1-(u+MiU/4)^2}} -\frac{1}{\sqrt{1-(u-MiU/4)^2}}$ 
			& $2\sum_{\alpha=\pm1}\sqrt{1-(u+\alpha MiU/4)^2} - MU$  \\ \hline 
			$M|w$                & $\mathbb{R}$         & $+1$       & $0$  & $-M$           & $0$                               & $0$   \\ \hline 	\end{tabular}
	\end{table*}

	First, for each  quasiparticle/string species $a$ we define a `particle' density that describes the distribution of occupied Bethe roots,
	\begin{equation}
	\!\rho_a(u) \simeq \frac{1}{L} \frac{\left(\text{\# of strings of species }a\text{ in }[u,u+du]\right)}{du}.
	\end{equation}
	We must also define distribution functions for {\it holes}. These can be understood as follows. Recall that for interacting integrable models, the allowed states  for a  string  depends on all the other strings in the system. Therefore, adding a string can generically displace the rapidities of  some of the other strings by an  $O(1)$ amount. For a given string species $a$, there is however a discrete set of rapidities $\left\{\bar{u}_j\right\}$ to which a string of species $a$ can be added, at the cost of displacing the other strings in rapidity space by only ${O}(1/L)$ (see e.g. Refs.~\onlinecite{takahashi_1999,essler_2005}). Again, in the thermodynamic limit, it is convenient to introduce hole densities that capture the distribution of the  $\left\{\bar{u}_j\right\}$,  viz.
	\begin{equation}
	\bar{\rho}_a(u) \simeq \frac{1}{L} \frac{\left(\text{\# of holes of species }a\text{ in }[u,u+du]\right)}{du}
	\end{equation}
	and hence the total densities,
	\begin{equation}
	\rho^t_a(u) = \rho_a(u) + \bar{\rho}_a(u).
	\end{equation}
	Note that each species has a specific set of rapidities that determine the domain of the corresponding distribution functions $\rho_a(u)$, $\bar{\rho}_a$, and $\rho^t_a$. The relevant spectral data are summarized in Table~\ref{tab:TBA_spectrum}.
	
	To complete the TBA description, we also require the scattering kernels $K_{ab}$ for every pair of string species, which are determined by the Hamiltonian of the model. If we consider the scattering of two strings of species $a$, $b$ and initial rapidities $u_a$, $u_b$, after their collision they will continue to  propagate in their initial direction with the same rapidities and species label, but they acquire a scattering phase shift: string $a$ acquires a phase shift $\phi_{ab}(u_a-u_b)$, and similarly,  string $b$ acquires a phase shift {$\phi_{ba}(u_b-u_a)$}. The scattering kernels are then defined via
	\begin{equation}
	K_{ab}(u) =  \frac{d}{du} \phi_{ab}(u) \equiv \phi'_{ab}(u)
	\end{equation}
	where in the second equation we have introduced  a notational convention whereby  derivatives with respect to the rapidity are denoted with a prime, that we adopt henceforth. The scattering kernels are summarized in Table~\ref{tab:TBA_scattering}, and depend on the pair of functions,
	\begin{align}
	K_M(u) &= \frac{1}{2\pi} \frac{UM}{2\left(u^2+M^2U^2/16\right)}\label{eq:KMNdef}\\
	K_{MN}(u) &= K_{M+N}(u) + K_{N-M}(u) + 2 \sum_{j=1}^{M-1} K_{N-M+2j}(u). \nonumber
	\end{align} 
	
	\begin{table}[]
		\caption{\label{tab:TBA_scattering} TBA Scattering kernels for the Hubbard model. Functions $K_M, K_{MN}$ are defined in Eqn.~\eqref{eq:KMNdef}.} 
		\begin{tabular}{|c|c|c|c|}
			\hline
			$\boldsymbol{K_{ab}}$                   & $\boldsymbol{y_\pm}$ & $\boldsymbol{N|w}$ & $\boldsymbol{N|uw}$ \\ \hline
			$\boldsymbol{y_\pm}$ & 0                    & $K_{N}$            & $K_{N}$             \\ \hline
			$\boldsymbol{M|w}$   & $K_{M}$              & $-K_{MN}$          & $0$            \\ \hline
			$\boldsymbol{M|uw}$  & $K_{N}$              & $0$           & $K_{MN}$           \\ \hline
		\end{tabular}
	\end{table}

	The quasiparticle densities and the scattering matrix together determine the admissible states, which are constrained to satisfy 
	\begin{equation}
	\label{eq:generic-density:app}
	\rho^t_a(u) = \left| \frac{1}{2\pi} k_a'(u) + \sum_b \left(K_{ab} \star \rho_b \right)(u)  \right|,
	\end{equation}
	where $k_a(u)$ denotes the quasi-momentum of strings of species $a$ with rapidity $u$	and $\star$ denotes the convolution as defined in the main text.
	For future convenience, we introduce $\sigma_a=\text{sign}(k_a'(u))$, so that we may rewrite \eqref{eq:generic-density:app} as
	\begin{equation}\label{eq:generic-density-mod:app}
	\sigma_a\rho^t_a(u) = \frac{1}{2\pi} k_a'(u) + \sum_b \left( K_{ab} \star \rho_b\right)(u).
	\end{equation}
	The set of constraint equations~\eqref{eq:generic-density-mod:app} are a transcription of the Bethe ansatz equations to the thermodynamic limit, and rewritten in terms of the appropriate distribution functions. This constraint, together with the maximum entropy principle, uniquely determine the distribution of energies of  the thermal generalized Gibbs state.~\cite{takahashi_1999,essler_2005,PhysRevB.96.081118} [{Note that there is a subtlety in systems with non-Abelian symmetries: the thermodynamic state obtained by applying the maximum entropy principle to the TBA spectrum only counts highest weight states in each multiplet. To see that this error is negligible,  let us consider a generic $L$-site system with an even number  $N$ of $SU(2)$ degrees of freedom each in the spin-$1/2$ representation, and assume that $N/L$ is held fixed as $N, L\to \infty$. The eigenspectrum can be decomposed into $SU(2)$ multiplets, with each representation $r\in\{0,1, 2,\ldots, N/2\}$ appearing $n_r$ times,  with energies $\epsilon_{r,j}$, where $j=1, 2,\ldots, n_r$. Then the {\it exact} free energy per site at inverse temperature $\beta=1/k_BT$ is given by $\beta f_{\text{ex}} = -\frac{1}{L} \ln\left[ \sum_{r=0}^{N/2} (2r+1) \sum_{j=0}^{n_r} e^{-\beta \epsilon_{r,j}}\right]$, whereas the TBA result  $\beta f_{\text{TBA}} = -\frac{1}{L} \ln\left[ \sum_{r=0}^{N/2} \sum_{j=0}^{n_r} e^{-\beta \epsilon_{r,j}}\right]$ neglects the degeneracy factors. However, the free energy difference per site, $\beta\delta f \equiv |\beta f_{\text{ex}} - \beta f_{\text{TBA}}| \leq \frac{\ln (N+1)}{L} $, which vanishes in the thermodynamic limit since $N$ is $O(L)$.}]

	Defining $Y_a(u)=\bar{\rho}_a(u)/\rho_a(u)$, the generalized Gibbs state can be shown to satisfy the {TBA equations}
	\begin{align}
	\label{eq:generic-TBA:app}
	\!\!\!\log Y_a(u)=\beta e_a(u)\! - \!\!\sum_b \left[K_{ab} \star \sigma_b \log\left(1+1/Y_b\right)\right]\!(u),\!
	\end{align}
	where $\beta$ is the inverse temperature and $e_a(u)$ is the (quasi)-energy of of strings of species $a$ with rapidity $u$, which is again determined by a microscopic energy-momentum relation derived from the underlying Hamiltonian.
	The set of functions $\{Y_a(u)\}$ completely characterize the state, and (as we will see below) are analogous to Boltzmann factors for the quasiparticles/strings. An equivalent set of functions is the set of filling factors, 
	\begin{equation}\label{eq:ffdef:app}
	n_a(u) = \rho_a(u)/\rho^t_a(u)= 1/[1+Y_a(u)].
	\end{equation}
	and the complementary hole filling factor $\bar{n}_a(u) = \bar{\rho}_a(u)/ \rho^t_a(u) = 1-n_a(u)$. The filling factors provide a convenient parameterization of a TBA state, via the (generically infinite-dimensional) vector $\bn = \{n_a(u)\}$. These play a central role in constructing a generalized hydrodynamic picture of integrable systems, where they are allowed to vary on long length and time scales. 
	
	\subsection{Dressing transformation}
	
	The bare momentum $k_a(u)$ and bare energy $e_a(u)$ defined above are necessary inputs for~\eqref{eq:generic-density:app} and ~\eqref{eq:generic-TBA:app}. However they 	 do not correspond to the physical momentum $k^{\text{ph}}$ (energy $e^{\text{ph}}$) of the string, measured as the momentum (energy) difference between the initial state, 	and the state with one added string of species $a$ and rapidity $u$. This is because, as mentioned already,  adding a single extra string to an unoccupied `hole'  causes a shift in the rapidities of all the other quasiparticles. Although each rapidity only shifts by $O(1/L)$, since a typical thermal state has  $O(L)$ such shifted quasiparticles, this `dressing'  gives an additional  $O(1)$ contribution to the physical momentum and  energy  of the resulting state. 
	Computing the `dressing' correction, the rapidity-derivatives of the physical quantities can be shown to satisfy the integral equations 
	\begin{align}
	\left(k_a^{\text{ph}}\right)'(u) &= \left(k_a'\right)^{\text{dr}}(u) =\left[\Omega_{ab} \star k_a' \right](u) \label{eq:kdressing:app}\\
	\left(e_a^{\text{ph}}\right)'(u) &= \left(e_a'\right)^{\text{dr}}(u) =\left[\Omega_{ab} \star e_a' \right](u), (u).\label{eq:edressing:app}
	\end{align}
	where we defined the dressing transformation
	\begin{align}
	f_a^\text{dr} &= \left[\Omega_{ab} \star f_b \right](u)
	\end{align}
	and dressing kernel $\Omega_{ab}$ as the unique inverse of the $\hat{1} - \hat{K} \sigma \hat{n}$ kernel, viz.
	\begin{equation}
	\sum_b\int \! dw\, \Omega_{ab}(u-w) \left[ \delta_{bc}\delta(w)  - K_{bc}(w)n_c(w)\right] = \delta_{ac} \delta(u)
	\end{equation}
	
	By relating the two different expressions for $k_a'(u)$ in~\eqref{eq:generic-density-mod:app} and \eqref{eq:kdressing:app}, we find that 
	\begin{align}\label{eq:rhoidentity:app}
	\rho^t_a(u) &=  \frac{1}{ 2\pi}\sigma_a \left(k^{\text{dr}}_a\right)'(u),
	\end{align}
	while applying a similar procedure to~\eqref{eq:generic-TBA:app} and \eqref{eq:edressing:app} and using \eqref{eq:ffdef:app} yields 
	\begin{align}\label{eq:Yidentity:app}
	Y_a(u)&=\exp\left[\beta e^{\text{ph}}_a(u)\right].
	\end{align}
	The identities \eqref{eq:ffdef:app} and \eqref{eq:Yidentity:app} have appealing physical interpretations: the first  indicates that the total density of states is obtained by  an appropriate derivative of the physical (dressed) momentum, and the second is consistent with the interpretation  of $Y_a$ as a generalized Boltzmann weight.

	In computing linear response, we also use the dressed charge $\mathcal{O}^{\text{dr}}$ corresponding to a conserved quantity $\mathcal{O}$. While {\it dressed} and {\it physical} quantities are formally similar, it is important to stress that they are different as the rapidity-derivative and the dressing transformation do not commute, e.g.
	\begin{equation}
	\left(e^{\text{ph}}\right)' = \left(e'\right)^{\text{dr}} \neq \left(e^{\text{dr}}\right)'
	\end{equation}
	Also conceptually, there is an important distinction between the two for globally conserved operators such as the electric charge and the magnetization. For the physical energy and the physical momentum, the rapidity shifts of the other quasiparticles alter the energy and momentum of the state with one added quasiparticle relative to the state when it is absent; this leads to a physical shift in the energy and momentum of the TBA eigenstate. In contrast, since $\hat{Q}$, $\hat{S}^z$ commute with ${H}$, such rapidity shifts cannot affect the physical {\it global} $U(1)$ charge of the TBA eigenstate, which simply changes by the bare (microscopic) value of the added quasiparticle. However, they can redistribute the local $U(1)$ charge between the quasiparticles and the background, which is captured by the `dressed' charge.  This is important when computing transport quantities or susceptibilities. An alternative way to compute the dressed charge that explains its physical meaning is as follows: Consider perturbing the Hamiltonian by an infinitesimal amount of a conserved charge $\mathcal{O}$, $H\mapsto H + \lambda_\mathcal{O} \mathcal{O}$,  and solving the TBA equations for the perturbed Hamiltonian; we then find that  $\mathcal{O}^{\text{dr}}_a = \frac{\partial}{\partial (\beta \lambda_\mathcal{O})}  \log Y_a$. Recalling that $\log Y_a = \beta e_a^{\text{ph}}$ we see that this measures the gradient in the energy of just that quasiparticle state with respect to chemical potential conjugate to $\mathcal{O}$, which intuitively corresponds to the dressed charge of $\mathcal{O}$ carried by the quasiparticle. However, the {\it global} charge in a given TBA eigenstate is simply $\langle {\mathcal{O}}\rangle$, as it should be.
	Note that when computing the energy response, the physical  and dressed energies are distinct, with the dressed energy being the quantity relevant to GHD response functions.

	\begin{table*}[]
		\caption{\label{tab:quasilocalTBA}Quasi-local form of the TBA equations for different functions $f_a(u)$ corresponding to TBA variables labeled by species and rapidity. Here $\mathcal{O}_a(u)$ corresponds to the projection of one of the two $U(1)$ conserved charges (the electric charge or the $z$-magnetization) onto quasiparticle species $a$ at rapidity $u$. $g_a(u)$ is meant to be either $e_a(u)$ or $k'_a(u)$.}
		\begin{tabular}{|c|rl|c|}
			\hline
			$\boldsymbol{f_a}$ & &   \textbf{TBA/Dressing Equations for} $\boldsymbol{f_a}$  & $\boldsymbol{\lim_{M\to\infty} \frac{f_a}{M}}$ \\
			\hline
			$\log Y_{\pm}$ & $\log Y_{\pm}=$ & $\beta \left( \tilde{e}_{\pm} - s \star \tilde{e}_{1|uw} \right) - s \star \log\left(\frac{1+Y_{1|w}}{1+Y_{1|uw}}\right)$ & ---\\
			$\log Y_{M|w}$ &  $\log Y_{M|w} =$ & $s \star I_{MN} \log\left(1+Y_{N|w}\right) - \delta_{M1} s \star \log\left(\frac{1+1/Y_-}{1+1/Y_+}\right)$ & $\beta h$\\
			$\log Y_{M|uw}$ & $\log Y_{M|uw}=  $ & $s \star I_{MN} \log\left(1+Y_{N|uw}\right) - \delta_{M1} s \star \log\left(\frac{1+Y_-}{1+Y_+}\right)$ & $-2\beta \mu$\\
			\hline
			$\mathcal{O}^{\text{dr}}_{\pm}$ & $\mathcal{O}^{\text{dr}}_{\pm} =$ &$ - s \star  \left[\bar{n}_{1|uw} \mathcal{O}^{\text{dr}}_{1|uw} - \bar{n}_{1|w} q^{\text{dr}}\right]$ & ---\\
			$\mathcal{O}^{\text{dr}}_{M|w}$ &$\mathcal{O}^{\text{dr}}_{M|w} = $ &$ s \star I_{MN} \bar{n}_{N|uw} \mathcal{O}^{\text{dr}}_{N|uw}
			- \delta_{M1} s \star \left[ n_- \mathcal{O}_{-}^{\text{dr}} -  n_+ \mathcal{O}_{+}^{\text{dr}} \right]$ &  $\lim_{M\to\infty} \frac{\mathcal{O}_{M|w}}{M}$\\
			$\mathcal{O}^{\text{dr}}_{M|uw}$ & $\mathcal{O}^{\text{dr}}_{M|uw} = $ & $s \star I_{MN} \bar{n}_{N|uw} \mathcal{O}^{\text{dr}}
			- \delta_{M1} s \star \left[ n_- \mathcal{O}_{-}^{\text{dr}} -  n_+ \mathcal{O}_{+}^{\text{dr}} \right]$ &  $\lim_{M\to\infty} \frac{\mathcal{O}_{M|uw}}{M}$\\
			\hline		
			$g^{\text{dr}}_{\pm}$ & $g^{\text{dr}}_{\pm} =$ & $ g_{\pm} - s\star g_{1|uw} - s \star  \left[\bar{n}_{1|uw} g^{\text{dr}}_{1|uw} - \bar{n}_{1|w} g^{\text{dr}}_{1|w}\right]$&---\\
			
			$g^{\text{dr}}_{M|w}$ & $g^{\text{dr}}_{M|w} = $ & $ s \star I_{MN} \bar{n}_{N|w} g^{\text{dr}}_{N|w}
			- \delta_{M1} s \star \left[ n_- g^{\text{dr}}_{-} -  n_+ g^{\text{dr}}_{+} \right]$ & $\lim_{M\to\infty} \frac{g_{M|w}}{M}$ \\
			
			$g^{\text{dr}}_{M|uw}$ & $g^{\text{dr}}_{M|uw} =$ & $ s \star I_{MN} \bar{n}_{N|uw} g^{\text{dr}}_{N|uw} 
			+ \delta_{M1} s \star \left[ \bar{n}_- g^{\text{dr}}_{-} -  \bar{n}_+ g^{\text{dr}}_{+} \right]$ & $\lim_{M\to\infty} \frac{g_{M|uw}}{M}$\\
			\hline
		\end{tabular}
	\end{table*}

	\section{Solving the TBA Equations\label{sec:app:TBAsol}}
	We now describe how to solve the TBA and dressing equations numerically in various regimes.	The generic form of the TBA equations~\eqref{eq:generic-TBA:app} and the various dressing equations~\eqref{eq:kdressing:app}, \eqref{eq:edressing:app} involve the kernels reported in Table~\ref{tab:TBA_scattering}. Given that $K_{MN}$ is non-zero for every pair $(M,N)$, the equations in this form contain a direct coupling among all magnon/singlet strings, making their interpretation and solution difficult. We refer to this as the ``non-local'' formulation of the TBA, in contrast with the one we are about to introduce. It is well known~\cite{essler_2005,takahashi_1999,PhysRevLett.121.230602} that there is an equivalent ``quasi-local'' formulation  where each string species is coupled to at most other $3$ string species. We now sketch the derivation of this quasilocal formulation of the TBA and the GHD dressing equations for the Hubbard model. Our notation closely parallels that of Ref.~\onlinecite{PhysRevB.96.081118}, who first derived the quasi-local form for the  dressing equations.
	
	We begin by defining the kernel $\left(\mathbf{1}+K\right)_{MN}^{-1}$ as the inverse under convolution of $\left(\mathbf{1}+K\right)_{MN}$, i.e.
	\begin{align}
	\left(\mathbf{1}+K\right)_{MN'}^{-1}\star \left(\mathbf{1}+K\right)_{N'M} = \mathbf{1}_{MN},
	\end{align}
	with $\mathbf{1}_{MN}(u)= \delta_{MN}\delta(u)$.
	Exploiting the explicit expression for $K_{MN}$, it can be shown that~\cite{takahashi_1999,essler_2005}
	\begin{equation}
	\left(\mathbf{1}+K\right)_{MN}^{-1}(u) =  \delta_{MN}\delta(u) - I_{MN} s(u),
	\end{equation}
	where we define \begin{align}
	s(u) &= [(\mathbf{\delta}+K_2)^{-1}\star K_1] (u)  = \frac{1}{U\cosh\left(2\pi u/U\right)}\\
	I_{MN} &=\delta_{M,N+1}+\delta_{M,N-1}.
	\end{align}
	From this, it is also easy to show that
	\begin{align}
	\left(\mathbf{1}+K\right)_{MN}^{-1}\star K_N = \left(\mathbf{1}+Is\right)_{MN}^{-1} \star K_N = \delta_{M1} s
	\end{align}
	
	Another property which will be useful in deriving the following equations is that for $f_M \in \{e_{M|uw}, k_{M|uw}'\}$, we have
	\begin{equation}
	\left(\mathbf{1}+K\right)_{MN}^{-1}\star f_M = \delta_{M1} s\star \left(f_+ - f_-\right),
	\end{equation}
	while for $g_M=\alpha M$ for any $\alpha$ we have
	\begin{align}
	\left(\mathbf{1}+K\right)_{MN}^{-1}\star g_M = 0
	\end{align}
	
	We now  act from the left with $\left(\mathbf{1}+K\right)_{MN}^{-1}$ on all terms in  the set of equations
	\begin{align}
	\log Y_{M|w} &= \beta e_{M|w} + K_{MN} \log\left(1+1/Y_{N|w}\right)\nonumber\\
	&~~ - K_M \star \log\left(\frac{1+1/Y_-}{1+1/Y_+}\right)\\
	\label{eq:non-local-TBA-uw}
	\log Y_{M|uw} &= \beta e_{M|uw} + K_{MN} \log\left(1+1/Y_{N|uw}\right)\nonumber\\
	&~~ - K_M \star \log\left(\frac{1+1/Y_-}{1+1/Y_+}\right)
	\end{align}
	and use the above identities to obtain quasi-local TBA equations for the singlets and the magnons  strings, listed in Table~\ref{tab:quasilocalTBA}.
	
	The case of the $y$-particles must be treated more carefully. Here, we first define $\boldsymbol{\tilde{e}}$ as the bare energy for $h=\mu=0$, and write  the TBA equation explicitly as
	\begin{align}
	\log Y_{\pm} &= \beta\left( \tilde{e}_{\pm} +\mu - h/2\right)\nonumber \\
	&~~+ K_M \star \log\left(\frac{1+1/Y_{M|uw}}{1+1/Y_{M|w}}\right).\label{eq:explicitTBAy}
	\end{align}
	We now first manipulate $K_M \star \log({1+1/Y_{M|w}})$ in order  to remove the coupling to all $M|w$ strings in favor of just the first, i.e. $1|w$.
	To do so, we fix a cutoff $\tilde{M}$, and for $M<\tilde M$ we rewrite
	\begin{align}
	\log(1+1/Y_{M|w}) &= \log(1+Y_{M|w})-\log Y_{M|w}\nonumber\\
	&= \log(1+Y_{M|w}) \nonumber\\
	&~~- I_{MN} s \star \log\left(1+Y_{N|w}\right)\nonumber\\
	&~~-\delta_{M1} s \star \log\left(\frac{1+1/Y_-}{1+1/Y_+}\right) 	\label{eq:non-loc-to-ql-y}
	\end{align}
	where we have used the quasi-local expression for $\log Y_{M|w}$ from Table~\ref{tab:quasilocalTBA}. We now substitute \eqref{eq:non-loc-to-ql-y} into $K_M \star \log({1+1/Y_{M|w}})$ in \eqref{eq:explicitTBAy}. We then manipulate the part coming from the first term in~\eqref{eq:non-loc-to-ql-y} as%
	\footnote{Note that the convolution product is associative.}
	\begin{align}
	K_M\star \log(1+Y_{M|w}) &= I_{MN} s \star K_N \star \log(1+Y_{M|w}) \nonumber
	\\&~~+ \delta_{M1} s \star \log(1+Y_{M|w})
	\end{align}
	In this way most of the terms cancel out and (after remembering to include the terms above the cutoff) we are left with
	\begin{align}
	&\sum_M K_M \star \log(1+1/Y_{M|w}) = \\
	&=\sum_{M>\tilde M} K_M \star \log(1+1/Y_{M|w})\nonumber\\
	&~~-K_1\star s\star \log\left(\frac{1+1/Y_-}{1+1/Y_+}\right)+s\star\log(1+Y_{1|w})\nonumber\\
	&~~-K_{\tilde{M}}\star s\star\log(1+Y_{\tilde{M}+1|w})+K_{\tilde{M}+1}\star s\star\log(1+Y_{\tilde{M}|w})\nonumber
	\end{align}
	Finally, using that $\log(1+1/Y_{M|w})=o(1/\revisionchange{M}^2)$ and that the last line in the previous expression converges to~\footnote{See for example the discussion in Ref.~\onlinecite{takahashi_1999} concerning the XXX chain.}
	\begin{align}
	\frac{1}{2} \lim_{\tilde{M}\to\infty}K_1\star\log(Y_{\tilde{M}|w})-\log(Y_{\tilde{M}+1|w})=-\frac{\beta h}{2},
	\end{align}
	we arrive at
	\begin{align}
	\label{eq:hybrid-TBA}
	\log Y_{\pm} &= \beta\left( \tilde{e}_{\pm} +\mu\right) \\
	&~~+ K_M \star \log\left(1+1/Y_{M|uw}\right) - s\star\log(1+Y_{1|w})\nonumber\\
	&~~+ K_1\star s\star \log\left(\frac{1+1/Y_-}{1+1/Y_+}\right).\nonumber
	\end{align}
	Finally, repeating the same procedure, but acting on $uw$-strings, we arrive at the quasi-local TBA equation for $y$-particles in Table~\ref{tab:quasilocalTBA}. The derivation for the quasi-local form the dressing equations closely follows those for the TBA equations.~\cite{PhysRevB.96.081118}
	
	To numerically solve the TBA equations, we truncate the hierarchy of $w$ strings ($uw$-strings) at some maximum length $\tilde{M}_w$ ($\tilde{M}_{uw}$). We introduce a rapidity cutoff for  the $w$ and $uw$ strings, requiring $u\in [-u_{\text{max}},u_{\text{max}}]$, and apporximate $Y_a(u)$ with $|u|>u_{\text{max}}$ as $Y_a\left(\text{sign}(u)u_{\text{max}}\right)$. We then discretize rapidity space into a regularly spaced grid containing $\tilde{m}_{\text{rap}}$ points per string. Finally, we solve the TBA and dressing equations iteratively. For example, focusing on the TBA equations, we take an initial guess $Y^{(0)}$ for $Y$. We plug it in the right-hand side of the TBA equations to compute the next approximation $Y^{(1)}$. We proceed in this way until $\left\lVert Y^{(j)}-Y^{(j-1)} \right\rVert_{\mathbb{L}^1}$ is less than the desired accuracy $\epsilon_{\text{acc}}$. We solve the equations for dressed and dressed quantities in a similar fashion (see also Ref.~\onlinecite{PhysRevB.101.035121}).
	
	While   truncating the hierarchy of strings,  in the non-local formulation we can just approximate all terms with the cutoff by $0$. However this truncation requires more care in the quasi-local formulation, where the asymptotic condition (the final column of Table~\ref{tab:quasilocalTBA}) is crucial to identify a unique solution. For the solution of the TBA equation, the issue is discussed in detail in Ref.~\onlinecite{PhysRevB.65.165104}. For the dressed quantities, we employ the following scheme. In order to compute  the dressing of  (for example) $f_{M|w}(u)$ with the asymptotic condition $\lim_{M\to\infty} f_{M|w}^{\text{dr}}/M=\alpha_w$,  we approximate
	\begin{equation}
	f_{\tilde{M}_w+1|w}^{\text{dr}}(u) \simeq K_1 \star f_{\tilde{M}_w|w}^{\text{dr}}(u) + \alpha_w.
	\end{equation} 
	
	Finally, we mention that in the spin-incoherent regime when $\beta\mu\lesssim-1$ and $\beta h\lesssim 1$, both formulations of the TBA equation presented above seem to be poorly suited to numerics. Instead, we find that the option that works best in this case is to employ a `hybrid' combination of Eq.~\eqref{eq:non-local-TBA-uw}, \eqref{eq:hybrid-TBA} and the quasi-local form for $w$-strings reported in Table~\ref{tab:TBA_scattering}.

	\section{Asymptotic results at large $M$ \label{sec:app:largeM}}
	
	In this section, building on Ref.~\onlinecite{PhysRevB.65.165104} and standard TBA results,~\cite{essler_2005,takahashi_1999} we expand the TBA equations at large string length $M$, with the goal of showing that
	\begin{align}
	\int du\, \rho_{M|w}^t(u) \left|v^\text{eff}_{M|w}(u)\right|& \sim \alpha/M^2\quad\text{ at }h=0\label{eq:limit:w}\\
	\int du\, \rho_{M|uw}^t(u) \left|v^\text{eff}_{M|uw}(u)\right|&\sim \tilde{\alpha}/M^2\quad\text{ at }\mu=0,\label{eq:limit:uw}
	\end{align}
	for two real numbers $\alpha$ and $\tilde{\alpha}$.
	For definiteness we focus on  ~\eqref{eq:limit:w}, but the argument proceeds identically for ~\eqref{eq:limit:uw}.
	
	First of all, using the definition of $v^{\text{eff}}$ we have
	\begin{equation}
	\int du\, \rho_{M|w}^t(u) \left|v^\text{eff}_{M|w}(u)\right| = \int \frac{du}{2\pi} \left|\left(e_{M|w}'\right)^{\text{dr}}\right|.\label{eq:limit:w:intermediate}
	\end{equation}
	Now, using the fact that that $e_{M|w}^{\text{ph}}(u)$ is even under $u\mapsto-u$ to halve the domain of integration, and the observation (see below) that $e_{M|w}(u)$ is monotonically increasing from $0$ to $+\infty$, we can finally re-express \eqref{eq:limit:w:intermediate}
	\begin{equation}
	\frac{1}{\pi\beta}\left( \log Y_{M|w}(+\infty)- \log Y_{M|w}(0)\right)
	\end{equation}
	
	Therefore we analyse the TBA equation for large ($M\gg1$) magnons in quasi-local form
	\begin{equation}
	\log Y_{M|w} = s\star \left[\log(1+Y_{M-1|w})+\log(1+Y_{M+1|w})\right]
	\end{equation}
	The solution of these equations at large $M$ and $h=0$ can be approximated by~\cite{PhysRevB.65.165104,PhysRevLett.121.230602}
	\begin{equation}
	\label{app:eq:Y_large_M}
	Y_{M|w}\simeq\left(f_{M}(u)+ M\right)^2-1,
	\end{equation}
	with $f_{M|w}=o(M)$ and $\lim_{u\to\infty}f_{M|w}(u)=0$. Plugging this form back into the TBA equation, we find that $f_M$ satisfies
	\begin{align}
	f_{M} &= s\star \left(f_{M-1}+f_{M+1}\right)\\
	\lim_{M\to\infty} f_M/M &=0,
	\end{align}
	where we have neglected terms of $o(f_M)$ and ${O}(1/M^4)$.
	One can verify that the recurrence relation can be rewritten as 
	\begin{equation}
	f_{M} = K_1 \star f_{M-1}.
	\end{equation}
	Thus $f_M \sim K_M \star \tilde{f}$ for some function $\tilde{f}$ with $\lim_{u\to\infty}\tilde{f}(u)=0$. As $M\to\infty$, we can neglect the width of $\tilde{f}$ relative to that of $K_M$. Therefore, the overall shape of $f_M$ is approximately given by a  Lorentzian with width proportional to $M$ and maximum height proportional to $1/M$.
	
	Finally, plugging this estimate into Eq.~\eqref{app:eq:Y_large_M} and expanding to leading order in $1/M$, we obtain the final result that $\log Y_{M|w}(+\infty)- \log Y_{M|w}(0)$ scales as $1/M^2$.
	
	\section{Large-$U$ expansions of the TBA \label{sec:app:largeU}}
	
	At large $U$, the TBA and dressing equations are considerably simplified, as was pointed out in Ref.~\onlinecite{takahashi_1999}. In this section we provide the complete solutions at leading order in $1/U$ in the strong-coupling regimes (i), (ii), and (iii), and a partial solution in regime (iv).
	The crucial observation which allows for exact solutions is that the kernels $s$ and $K_M$ have a width of order $U$ and an height of order $1/U$, and that $y$-particle rapidities are bounded in the $[-1,1]$ interval. We observe that for a function $f$ with domain $[-1,1]$, we can expand
	\begin{align}
	\label{app:eq:y-convolution-expansion}
	\left(s\star f\right)(u) &= \left(\int_{-1}^{+1}dw\,f(w)\right) s(u) \\
	&~~- \left(\int_{-1}^{+1}dw\,w f(w)\right) s'(u) + {O}(1/U^3). \nonumber
	\end{align}
	On the other hand, for a function $g$ with domain in $\mathbb{R}$, we can expand
	\begin{align}
	\label{app:eq:string-convolution-expansion}
	\left.\left(s\star f\right)\right\vert_{[-1,1]}(u) &= \int dw\, K(-w) g(w) \\ 
	&~~+u\int dw\,  K'(-w) g(w)+ {O}(|g|/U^2).\nonumber
	\end{align}
	Note that the same expansion holds also if $s$ is replaced by $K_M$.
	
	\subsection{Regimes (i), (ii) and (iii): $U\gg1\gg\beta$}
	
	In this case, it is convenient to work with the quasi-local formulation of the TBA and dressing equations.
	From Eq.~\eqref{app:eq:y-convolution-expansion}, we see that
	\begin{equation}
	\log Y_{\pm}(u) = \beta e_{\pm}(u) + \alpha_Y + {O}(1/U^3),
	\end{equation}
	where $\alpha_Y$ is $u$-independent.
	Furthermore the strings satisfy the equations
	\begin{align}
	\label{app:eq:iv-TBA-w}
	\log Y_{M|w} &= s \star I_{MN} \log\left(1+Y_{N|w}\right) + \delta_{M1} \gamma_Y s\\
	\label{app:eq:iv-TBA-uw}
	\log Y_{M|uw} &= s \star I_{MN} \log\left(1+Y_{N|uw}\right) + \delta_{M1} \tilde{\gamma}_Y s\\
	\gamma_Y&= -\int_{-1}^{+1} \log \left(\frac{1+1/Y_{-}}{1+1/Y_{+}}\right)\\
	\tilde{\gamma}_Y &= -\int_{-1}^{+1} \log \left(\frac{1+Y_{-}}{1+Y_{+}}\right).
	\end{align}
	The key observation is that $\left\Vert s \right\Vert_{\infty}={O}(1/U)$, so that at leading order we can just neglect the terms $\gamma_Y s$ and $\tilde{\gamma}_Y s$, in which case~\eqref{app:eq:iv-TBA-w} and ~\eqref{app:eq:iv-TBA-uw} are simply the $T=\infty$ TBA equations for the strings. Their solution is known to be given by~\cite{takahashi_1999}
	\begin{align}
	Y_{M|w} &= \chi_M^2 - 1, &\chi_M = \frac{\sinh\left[(M+1)\beta h /2\right]}{\sinh\left(\beta h/2\right)}\\
	Y_{M|uw} &= \tilde{\chi}_M^2 - 1, &\tilde{\chi}_M = \frac{\sinh\left[(M+1)\beta \mu\right]}{\sinh\left(\beta \mu\right)}.
	\end{align}
	{Plugging $Y_{1|w}$ and $Y_{1|uw}$ into the equation for $\log Y_{\pm}$, we can now determine}
	\begin{equation}
	\alpha_Y = \log \tilde{\chi}_1 - \log \chi_1,
	\end{equation}
	thus completing the solution of the TBA equation.
	
	Nonetheless, it will be convenient compute dressed charges using 
	\begin{align}
	\tilde{m}^{\text{dr}}&=\frac{1}{\beta}\frac{\partial\log Y}{\partial h},\\
	\tilde{q}^{\text{dr}}&=\frac{1}{\beta}\frac{\partial\log Y}{\partial\mu},\\
	\tilde{e}^{\text{dr}}&=\left.\frac{\partial\log Y}{\partial\beta}\right\vert_{\beta\mu,\beta h},
	\end{align}
	from $Y$-functions and $(e')^{\text{dr}} = \left(e^{\text{ph}}\right)'$ from $Y_a(u) = \exp[\beta e^{\text{ph}}_a(u)]$ and .
	Then to obtain a non-zero result we need to expand to next order and take into account the source term $\gamma_Y s$ and $\tilde{\gamma}_Y s$.
	We then expand $\log Y = \beta e^{\text{ph}} = \log Y^{(0)} + \beta \varepsilon^{(1)} + {O}(1/U^2)$ with  $\varepsilon^{(1)}={O}(1/U)$. Plugging this ansatz into the Eq.~\eqref{app:eq:iv-TBA-w}~\eqref{app:eq:iv-TBA-uw}, we obtain
	\begin{align}
	\label{app:eq:iv-e_dr_w}
	\varepsilon^{(1)}_{M|w} &= s \star I_{MN} \left(1-n_{N|w}\right) \varepsilon^{(1)}_{N|w} + \delta_{M1} \frac{\gamma_Y}{\beta} s,\\
	\varepsilon^{(1)}_{M|uw} &= s \star I_{MN} \left(1-n_{N|uw}\right) \varepsilon^{(1)}_{N|uw} + \delta_{M1} \frac{\tilde{\gamma}_Y}{\beta} s,
	\end{align}
	where we can substitute $n_a$ with its approximate form to leading order in $1/U$,
	\begin{equation}
	n_a = \frac{1}{1+\exp(\log Y^{(0)}_a)}.
	\end{equation}
	This set of equations has been solved exactly~\cite{PhysRevLett.121.230602} and yields
	\begin{align}
	\varepsilon^{(1)}_{M|w} &= \gamma_Y \frac{\chi_M}{\chi_1} \left( \frac{K_M}{\chi_{M-1}} - \frac{K_{M+1}}{\chi_{M+1}} \right),\\
	\varepsilon^{(1)}_{M|uw} &= \tilde{\gamma}_Y \frac{\tilde{\chi}_M}{\tilde{\chi}_1} \left( \frac{K_M}{\tilde{\chi}_{M-1}} - \frac{K_{M+1}}{\tilde{\chi}_{M+1}} \right).
	\end{align}
	This directly allows for the computation of dressed charges by taking derivatives.
	
	The last quantity left to compute is $(k^{\text{dr}})'$, which also yields $\rho^t$. This can be solved in a similar fashion to the TBA equations, yielding
	\begin{align}
	\left(k'_{\pm}\right)^{\text{dr}} &= k'_{\pm},\\
	\left(k'_{M|w}\right)^{\text{dr}} &= \gamma_k \frac{\chi_M}{\chi_1} \left( \frac{K_M}{\chi_{M-1}} - \frac{K_{M+1}}{\chi_{M+1}} \right),\\
	\left(k'_{M|uw}\right)^{\text{dr}} &= \tilde{\gamma}_k \frac{\tilde\chi_M}{\tilde\chi_1} \left( \frac{K_M}{\tilde\chi_{M-1}} - \frac{K_{M+1}}{\tilde\chi_{M+1}} \right),\\
	\gamma_k &= \int_{-1}^{+1} du\, \left(n_- k'_- - n_+ k'_+\right),\\
	\tilde{\gamma}_k &= \int_{-1}^{+1} du\, \left[(1-n_-) k'_- - (1-n_+) k'_+\right].
	\end{align}
	
	{
		Note that in practice this expansion is valid across regimes (i-iii), although conceptually regimes (ii-iii) can be accessed more easily using a hybrid form (see Appendix~\ref{sec:app:TBAsol}), where $uw$-strings explicitly drop out of the problem. While this does not affect the analytical solution, it is extremely convenient numerically, as it allows use to retain a very low cutoff on the length of $uw$-strings and hence improves the convergence of the iterative solution.
		
		From the data above we can explicitly compute Drude weights and correlators. The Drude weights are always dominated by $y$-particles since 
		\begin{equation}
		\int du\, \rho_{M|w}^t(u) \left(v^\text{eff}_{M|w}(u)\right)^2={O}(1/U^2).
		\end{equation}
		and similarly for $uw$-strings.
		Furthermore, the contribution of $w$-strings  energy and charge correlators is again suppressed since
		\begin{equation}
		\left(q^{\text{dr}}_{M|w}(u)\right)^2 = {O}(1/U^2),
		\end{equation}
		and similarly for the component of the dressed energy from $\tilde{e}$ (Table~\ref{tab:TBA_spectrum}).
		Finally, as shown above the dressed quantities for $y$-particles are unaffected by the dressing to leading order in $1/U$.
		
		As mentioned in the main text, the expression Eq.~\eqref{eq:large-U-Drude} for the Drude weight can be further simplified in regime (iii), where $\beta n_{\pm}(1-n_{\pm})$ is significantly non-zero only over an interval of width $T$ around the fermi points $u_F$ where $e_\pm(u_F)=0$.
		Linearising $e_\pm(u_F)=0$ around the Fermi points, and explicitly performing the integral over rapidity we obtain the low-temperature form of the Drude weight in Eq.~\eqref{eq:large-U-low-T-Drude}.

		We now proceed to estimate the order of the corrections to the Drude weights in regime (iii). The main source of these corrections is given by the feedback of $\varepsilon^{(1)}$ and $\left(k'\right)^\text{dr}$ of $w$-strings (and $uw$-strings in regime (i)) in the TBA equation and dressing equations. For definiteness we focus on $\left(k'\right)^\text{dr}$, but the reasoning is the same for $\varepsilon^{(1)}$ in the TBA equations. Looking at Eq.~\ref{app:eq:string-convolution-expansion}, and noticing that $\left(k'_{M|uw}\right)^{\text{dr}}(u)$ is even under $u\mapsto-u$, we conclude that the leading correction is constant in $u$ and is of order ${O}(1/U)$. A similar consideration applies to $e^{\text{ph}}_\pm$.
		The corrections we obtain at order $1/U$ can be obtained by making  the following replacements in Eq.~\eqref{eq:large-U-low-T-Drude}
		\begin{align}
		e_{\pm} &\to e_\pm + \gamma_Y \int du\,  \left( \frac{K_1(u)}{\chi_{0}} - \frac{K_{2}(u)}{\chi_{2}} \right) s(u) \\
		&~~-\tilde{\gamma}_Y \int du\, s(u)K_1(-u) ,\nonumber\\
		k_{\pm}' &\to k_\pm' + \gamma_k \int du\,  \left( \frac{K_1(u)}{\chi_{0}} - \frac{K_{2}(u)}{\chi_{2}} \right)s(u) \\
		&~~-\tilde{\gamma}_k \int du\, s(u)K_1(-u),\nonumber
		\end{align}
		and modifying the Fermi velocity and the dressed charges accordingly. As anticipated, the corrections are $O(1/U)$, due to
		\begin{equation}
		\int du\, K_M(u)s(u) = {O}(1/U)
		\end{equation}
		
		Finally, we focus on regimes (ii) and (iii) at $\mu=h=0$. In this case the contribution of $y$-particles to the Drude weight is exponentially suppressed by $\beta U$. Therefore, we focus on the string  contributions to the energy Drude weight, and find that
		\begin{align}
		D_{M|w} &= \frac{\gamma_Y^2 \left(\partial_\beta \gamma_Y\right)^2}{\beta \gamma_k} T(M) n_{M|w}(1-n_{M|w}), \\
		D_{M|uw} &= \frac{\tilde{\gamma}_Y^2 \left(\partial_\beta\tilde{\gamma}_Y\right)^2}{\beta \tilde{\gamma}_k} T(M) n_{M|uw}(1-n_{M|uw}) ,
		\end{align}
		with
		\begin{widetext}
			\begin{equation}
			T(M) = \frac{40+540 M+3198 M^2+10901 M^3+23472 M^4+32562 M^5+28274 M^6+14016 M^7+3040 M^8}{256 M^7 (1+M) (2+M)^3 (1+2 M)^4 \pi ^2 (U/4)^4}.
			\end{equation}
		\end{widetext}
		It then remains to evaluate the factor multiplying $T(M)$. To do this, note that $\log Y_{\pm} =- \beta U/2+{O}(1)$, meaning that $n_\pm=1$, with $1-n_\pm$ exponentially suppressed in $\beta U/2$. From this, we can notice that $D_{M|uw}$ is exponentially suppressed, furthermore $\gamma_Y/\beta \simeq \gamma_k \simeq 2\pi$, obtaining Eq.~\eqref{eq:energyDrudehalf-explicit}.

		\subsection{Regime (iv): $\beta\gg U\gg1$}
		
		In this regime, we deploy the standard technology of $T\to 0$ TBA expansions.~\cite{takahashi_1999,essler_2005,PhysRevB.9.2150,PhysRevB.99.014305,Bertini_2018} The fundamental idea behind the simplification of the TBA equations in this limit is to express them in terms of dressed energy $e^{\text{ph}}= \log Y/\beta$ which remain finite as $\beta \to \infty$. In non-local form, the equations for the dressed energies then become (at leading order in $\beta^{-1}$)
		\begin{equation}
		\label{eq:app:zero-T-TBA}
		e^{\text{ph}}_a = e_a - K_{ab}\star \left[e^{\text{ph}}_b \right]^-,
		\end{equation} 
		with $[f]^-=f\theta_H(-f)$, denoting the Heaviside-$\theta$ function with $\theta_H$. Furthermore $n_a=\theta_H(-e^{\text{ph}}_a)$.
		
		The first consequence is that singlets ($uw$-strings) always have~\cite{takahashi_1999,essler_2005} $n_{M|uw}=0$. Specifically in the high-$U$ limit, from Eq.~\eqref{app:eq:string-convolution-expansion} it follows again that energy, momentum and charge of $y$-particles are dominated by bare quantities. On the other hand, Eq.~\eqref{app:eq:y-convolution-expansion} implies that $e^{\text{ph}}_{M|w}(u)$ are functions with height ${O}(1/U)$ and rapidity-width ${O}(U)$, and similar considerations hold for $(k'_{M|w})^{\text{dr}}(u)$. Thus, the velocities of $w$-strings are again suppressed by factors of $1/U$, implying that  in regime (iv) as well the Drude weight is dominated by $y$ particles. These same considerations lead us also to conclude that the dressed charge and energy of $y$ particles are dominated by their bare value. Therefore, the crossover between regimes (iii) and (iv) cannot be observed in the (reduced) energy or charge Drude weights. However, the magnetization of $y$-particles is significantly corrected relative to its bare value due to scattering off the magnons, and the spin Drude weight picks up the crossover due to the change in magnon properties.
		
		Although a  full analytical solution in this regime is generally hard,~\cite{essler_2005} we focus on the $h\ll t$ regime to show that the leading corrections to the charge and energy Dude weight in Eq.~\eqref{eq:large-U-low-T-Drude} are different than those in regime (iii).
		When $h\ll t$, the dressed energy of the ``elementary'' magnon (i.e. the $1|w$ string) is negative in a large rapidity interval centred around $0$. Approximating this interval by the whole real axis, the TBA equation becomes linear in $e^{\text{ph}}_{1|w}$, and furthermore all higher magnons ($M|w$ strings with $M>1$) strings drop out of the problem. The TBA equation can then be readily solved (to leading order in $1/U$)
		\begin{align}
		e^{\text{dr}}_\pm &= e_\pm,\\
		e^{\text{dr}}_{1|w} &= \gamma_Y s(u)<0.
		\end{align}
		The dressing equation for $\left(k^{\text{dr}}\right)'$ yields a similar result
		\begin{align}
		\left(k'_\pm\right)^{\text{dr}} &= e_\pm,\\
		\left(k'_{1|w}\right)^{\text{dr}} &= \gamma_k s(u)<0.
		\end{align}
		The correction to the Drude weight can then be computed similarly to that in case (iii), by the following substitutions in Eq.~\eqref{eq:large-U-low-T-Drude}
		\begin{align}
		e_{\pm} &\to e_\pm + \gamma_Y \int du\, K_1(u) s(u), \\
		k_{\pm}' &\to k_\pm' + \gamma_k \int du\, K_1(u) s(u),
		\end{align}
		where, again
		\begin{equation}
		\int du\, K_1(u) s(u) = {O}(1/U).
		\end{equation}

		Finally, we discuss energy transport in regime (iv) at $\mu=h=0$. In particular, we will show that $D_e\sim\alpha_e T^2$ as $T\to0$ and we will compute $\alpha_e$ to leading order in $1/U$. We use the fact that at $\mu=h=0$, in the limit $T\to0$, the solution of the zero-temperature TBA equations~\eqref{eq:app:zero-T-TBA}, is known~\cite{essler_2005,PhysRevB.9.2150} and in particular
		\begin{align}
		e^{\text{ph}}_\pm&<0\\
		e^{\text{ph}}_{1|w}&=s\star\left(e_-^{\text{ph}}-e_+^{\text{ph}}\right)_-<0\\
		e^{\text{ph}}_{M|w}&=0 \text{ for }M>1\\
		e^{\text{ph}}_{M|uw}&=e_{M|uw} - K_M\star\left(e_-^{\text{dr}}-e_+^{\text{dr}}\right)_- >0\\
		\lim_{u\to\pm\infty}&e^{\text{ph}}_{1|w}(u)=0\\
		\lim_{u\to\pm\infty}&e^{\text{ph}}_{M|uw}(u)=0.
		\end{align}
		Finally, the dressed momenta will satisfy
		\begin{align}
		\left(k_{1|w}'\right)^{\text{dr}}&=s\star\left[n_-\left(k_-^{\text{dr}}\right)'-n_+\left( k_+^{\text{dr}}\right)'\right]_-<0\\
		\left(k_{M|w}'\right)^{\text{dr}}&=0 \text{ for }M>1\\
		\left(k_{M|uw}'\right)^{\text{dr}}&=k_{M|uw}' \\
		&~~~-K_M\star\left[n_-\left(k_-'\right)^{\text{dr}}-n_+\left( k_+'\right)^{\text{dr}}\right]_- >0,\nonumber\\
		\end{align}
with the limiting behavior
\begin{align}
		\lim_{u\to\pm\infty}&\left(k'_{1|w}\right)^{\text{dr}}(u)&=0\\
		\lim_{u\to\pm\infty}&\left(k'_{M|uw}\right)^{\text{dr}}(u)&=0.
		\end{align}
		
		As pointed out in the previous section,transport in the $T\to0$ limit is dominated by the Fermi points. The contribution to the energy Drude weight from a species $a$ will be
		\begin{align}
		D_{e,a} \!=\!  \beta \int\!\! du \left(e^{\text{ph}}_a(u)\right)^2 \left(v^{\rm eff}_a(u)\right)^2 \rho^t_a(u) (1-n_a(u)) n_a(u),
		\end{align}
		where we used the fact  $e^{\text{ph}}=e^{\text{dr}}+O(T^2)$, which follows from the low-T expansion in Ref.~\onlinecite{PhysRevB.99.014305}.
		The contribution of each Fermi point can be computed by linearising $e^{\text{ph}}$ near the Fermi point. If the Fermi point is at rapidity $u_F$ for the species $a$, its contribution is  
		\begin{align}
		\frac{v^{\text{eff}}_{a}(u_F) T^2}{2\pi} \int_{-\infty}^{+\infty} dy\, y^2 \frac{e^{y}}{\left(1+e^y\right)^2} = \frac{v^{\text{eff}}_a(u_F) T^2 \pi}{6}.
		\end{align}
		
		{In the present case,  however, $M|uw$ and $1|w$-strings lack Fermi points at finite rapidities, but their physical energies $e^{\text{ph}}(u)\to0$ as $u\to\infty$; we may therefore think of these strings as having Fermi points shifted to infinite rapidity.} A similar calculation then shows that
		\begin{align}
		D_{e,a} = \frac{v^{\text{eff}}_a(\infty) T^2 \pi}{6},
		\end{align}
		where we assumed that $\lim_{u\to\pm\infty} v^{\text{eff}}_a(u) = v^{\text{eff}}_a(\infty)$ is finite.
		
		Before computing $v^{\text{eff}}(\infty)$, we discuss the contribution of $M|w$-strings for $M>2$. Both their dressed energy and their total density will be~\cite{PhysRevB.99.014305} $O(T^2)$. From the behavior of the physical energy, we have that the dressed energy is  $O(T^3)$. Therefore the contribution to the thermal Drude weight from $M|w$-strings will be $O(T^6)$, which is negligible in comparison to the Fermi point contributions.
		
		We are now left with computing $v^{\text{eff}}(\infty)$ to leading order in $1/U$. We start by focusing on, $v^{\text{eff}}_{1|w}(u)$ and exploit the expansion~\eqref{app:eq:y-convolution-expansion} to obtain
		\begin{equation}
		v^{\text{eff}}_{1|w}(\infty) = \frac{\gamma_Y}{\gamma_k} \lim_{u\to\infty}\frac{s'(u)}{s(u)} = \frac{\gamma_Y}{\gamma_k} \frac{2\pi}{U}.
		\end{equation}
		On the other hand, for $uw$-strings both $e^{\text{ph}}_{M|uw}(u)$ and $\left(k'_{M|uw}\right)^{\text{dr}}(u)$ decay as $1/u^2$ at large rapidity $u$. Thus $v^{\text{eff}}_{M|uw}(\infty)=0$ and $uw$-strings again do not contribute to transport.
		
		Combining the above results, we see that
		\begin{equation}
		D_e = \frac{\pi^2 T^2}{3U}+O(T^4) +O(T^2/U^2),
		\end{equation}
		or, reintroducing the $t$-dependence through dimensional analysis, $D_e\simeq \frac{\pi^2 t^3 T^2}{3U}$.
	}

\end{appendix}

\end{document}